\documentclass[longauth]{aa}

\usepackage[utf8]{inputenc}
\usepackage{graphicx}

\usepackage{amsmath}
\usepackage{float}
\usepackage{natbib}
\bibpunct{(}{)}{;}{a}{}{,} 
\usepackage{array}
\usepackage{hyperref}
\usepackage{subfig}
\usepackage[varg]{txfonts}
\usepackage{multirow}
\usepackage{siunitx}
\usepackage{upgreek}
\usepackage{textgreek}
\usepackage{ulem}
\usepackage{titlesec}
\usepackage{tablefootnote}
\usepackage{lscape}
\usepackage{xcolor}
\usepackage{makecell}
\usepackage{adjustbox}

\hypersetup{colorlinks=true,citecolor=blue}

\titleformat{\paragraph}
{\normalfont\normalsize\bfseries}{\theparagraph}{1em}{}
\titlespacing*{\paragraph}
{0pt}{3.25ex plus 1ex minus .2ex}{1.5ex plus .2ex}

\def\app#1#2{%
  \mathrel{%
    \setbox0=\hbox{$#1\sim$}%
    \setbox2=\hbox{%
      \rlap{\hbox{$#1\propto$}}%
      \lower1.1\ht0\box0%
    }%
    \raise0.25\ht2\box2%
  }%
}

\DeclareTextSymbol{\degre}{T1}{6}
\DeclareTextSymbol{\degre}{OT1}{23}

\makeatletter
\g@addto@macro{\endtabular}{\rowfont{}}
\makeatother
\newcommand{\rowfonttype}{}
\newcommand{\rowfont}[1]{
\gdef\rowfonttype{#1}#1\ignorespaces%
}
\makeatother

\defcitealias{}{}          

\begin{document}
  \title{Homogeneous search for helium in the atmosphere of 11 gas giant exoplanets with SPIRou }

   \author{ R. Allart\inst{1,*,\thanks{Trottier Postdoctoral Fellow}} ,
   P.-B. Lem\'ee-Joliecoeur\inst{1},
   A. Y. Jaziri\inst{2},
   D. Lafreni\`ere\inst{1},
   E. Artigau\inst{1},
   N. Cook\inst{1},
   A. Darveau-Bernier\inst{1},
   L. Dang\inst{1,\thanks{Banting Postdoctoral Fellow}}
   C. Cadieux\inst{1},
   A. Boucher\inst{1},
   V. Bourrier\inst{2},
   E. K.~Deibert\inst{3},
   S. Pelletier\inst{1},
   M. Radica\inst{1}, 
   B. Benneke\inst{1},
   A. Carmona\inst{4},
   R. Cloutier\inst{5},
   N. B. Cowan\inst{6,7},
   X. Delfosse\inst{4},
   J.-F. Donati\inst{8},
   R. Doyon\inst{1},
   P. Figueira\inst{2},
   T. Forveille\inst{4},
   P. Fouqu\'e\inst{8},
   E. Gaidos\inst{9},
   P.-G. Gu\inst{10},
   G. H\'ebrard\inst{11,12},
   F. Kiefer\inst{13},
   \'A K\'osp\'al\inst{14,15,16,17},
   R. Jayawardhana\inst{18},
   E. Martioli\inst{19,11},
   L. A. Dos Santos\inst{20}, 
   H. Shang\inst{21},
   J. D. Turner\inst{22}\thanks{NHFP Sagan Fellow}, and
   A. Vidotto\inst{23}
}
   
   \institute{
   \inst{1} D\'epartement de Physique, Institut Trottier de Recherche sur les Exoplan\`etes, Universit\'e de Montr\'eal, Montr\'eal, Qu\'ebec, H3T 1J4, Canada\\
   \inst{2} Observatoire astronomique de l'Universit\'e de Gen\`eve, Universit\'e de Gen\`eve, chemin Pegasi 51, CH-1290 Versoix, Switzerland\\
   \inst{3} Gemini Observatory, NSF's NOIRLab, Casilla 603, La Serena, Chile\\
   \inst{4} Universit\'e Grenoble Alpes, CNRS, IPAG, 38000 Grenoble, France \\
   \inst{5} Dept. of Physics \& Astronomy, McMaster University, 1280 Main St West, Hamilton, ON, L8S 4L8, Canada \\
   \inst{6} Department of Physics, McGill University, 3600 rue University, Montr\'eal, QC, H3A 2T8, Canada\\
   \inst{7} Department of Earth \& Planetary Sciences, McGill University, 3450 rue University, Montr\'eal, QC, H3A 0E8, Canada\\
   \inst{8} Institut de Recherche en Astrophysique et Plan\'{e}tologie, Universit\'{e} de Toulouse, CNRS, 14 avenue Edouard Belin, F-31400, Toulouse, France\\
   \inst{9}  Department of Earth Sciences, University of Hawai\'i at Manoa, Honolulu, HI 96822 USA\\
   \inst{10} Institute of Astronomy and Astrophysics, Academia Sinica, Taipei 10617, Taiwan\\
   \inst{11} Institut d'Astrophysique de Paris, CNRS, UMR 7095, Sorbonne Universit\'{e}, 98 bis bd Arago, 75014 Paris, France\\
   \inst{12} Observatoire de Haute Provence, St Michel l’Observatoire, France\\
   \inst{13} LESIA, Observatoire de Paris, Universit\'e PSL, CNRS, Sorbonne Universit\'e, Universit\'e Paris Cit\'e, 5 place Jules Janssen, 92195 Meudon, France\\
   \inst{14} Konkoly Observatory, Research Centre for Astronomy and Earth Sciences, E\"otv\"os Lor\'and Research Network (ELKH), Konkoly-Thege Mikl\'os \'ut 15-17, 1121 Budapest, Hungary\\
   \inst{15} CSFK, MTA Centre of Excellence, Konkoly-Thege Mikl\'os \'ut 15-17, 1121 Budapest, Hungary\\
   \inst{16} ELTE E\"otv\"os Lor\'and University, Institute of Physics, P\'azm\'any P\'eter s\'et\'any 1/A, 1117 Budapest, Hungary\\
   \inst{17} Max Planck Institute for Astronomy, K\"onigstuhl 17, 69117 Heidelberg, Germany\\
   \inst{18} Department of Astronomy, Cornell University, Ithaca, NY 14853, U.S.A.\\
   \inst{19} Laborat\'orio Nacional de Astrof\'isica, Rua Estados Unidos 154, Itajub\'a, MG 37504364, Brazil\\
   \inst{20} Space Telescope Science Institute, 3700 San Martin Drive, Baltimore, MD 21218, USA\\
   \inst{21} Institute of Astronomy and Astrophysics, Academia Sinica,  Taipei 10617, Taiwan\\
   \inst{22} Department of Astronomy and Carl Sagan Institute, Cornell University, 122 Sciences Drive, Ithaca, NY 14853, USA\\
   \inst{23} Leiden Observatory, Leiden University, PO Box 9513, 2300 RA Leiden, The Netherlands\\
              		* \email{romain.allart@umontreal.ca}
             }

   \date{Received January 1, 2015; accepted January 1, 2015}

\abstract
{The metastable helium triplet in the near-infrared (10833\,\AA) is among the most important probes of exoplanet atmospheres. It can trace their extended outer layers and constrain mass-loss. We use the near-infrared high-resolution spectropolarimeter SPIRou on the CFHT to search for the spectrally resolved helium triplet in the atmospheres of eleven exoplanets, ranging from warm mini-Neptunes to hot Jupiters and orbiting G, K and M dwarfs. Observations were obtained as part of the SPIRou Legacy Survey and complementary open-time programs. We apply a homogeneous data reduction to all datasets and set constraints on the presence of metastable helium, despite the presence of systematics in the data. We confirm published detections for HAT-P-11\,b, HD\,189733\,b, and WASP-69\,b and set upper limits for the other planets. We apply the \texttt{p-winds} open source code to set upper limits on the mass-loss rate for the non-detections and to constrain the thermosphere temperature, mass-loss rate, line-of-sight velocity, and the altitude of the thermosphere for the detections. We confirm that the presence of metastable helium correlates with the stellar mass and the XUV flux received by the planets. We investigated the correlation between the mass-loss rate and the presence of metastable helium, but it remains difficult to draw definitive conclusions. Finally, some of our results are in contradiction with previous results in the literature, therefore we stress the importance of repeatable, homogeneous, and larger-scale analyses of the helium triplet to obtain robust statistics, study temporal variability, and better understand how the helium triplet can be used to explore the evolution of exoplanets.}

   \keywords{Planets and satellites: atmospheres, planets and satellites: gaseous planets, infrared: planetary systems, instrumentation: spectrographs, techniques: spectroscopic, methods: observational}
   \titlerunning{SPIRou helium survey}
   \authorrunning{R. Allart, P.-B. Lem\'ee-Joliecoeur, Y. Jaziri et al. }
   \maketitle

\section{Introduction}
Through their lifetime, exoplanets might undergo several physical processes that can alter their compositions, masses and sizes. Being the outer envelope of exoplanets, atmospheres are excellent windows onto the exoplanets and particularly subject to their evolution (be it impinging radiation, mass loss, etc). The atmospheres of close-in gas giant planets hydrodynamically expand under the absorption of stellar irradiation \citep[e.g.,][]{vidal-madjar_extended_2003, lammer_atmospheric_2003} and, in extreme conditions, they can evaporate and be stripped away from the planet core \citep{bourrier_hubble_2018,owen_photoevaporation_2018}. Such atmospheric changes could happen in the early stage of the system (100\,Myr-1\,Gyr, \cite{owen_atmospheric_2019,johnstone_young_2021,king_euv_2021}), in particular, once the gaseous protoplanetary disk has been dissipated and the planet is directly irradiated by the central star for planets either formed in-situ \citep{batygin_situ_2016,matsumoto_size_2021} or during their disk-driven inward migrations \citep[e.g.,][]{ida_toward_2008}. Under such intense irradiation, Neptune-sized planets could be unable to retain their gaseous envelopes and could even become bare cores due to their lower initial mass. This is consistent with the observed lack of close-in hot Neptunes (Fig.\,\ref{fig:Population}), commonly called the Neptunian or the evaporation desert \citep[e.g.,][]{lecavelier_des_etangs_diagram_2007,beauge_emerging_2013,mazeh_dearth_2016,owen_atmospheric_2019}. Another explanation for the lack of hot Neptunes can be found in high-eccentricity migration scenario \citep[e.g.,][]{rasio_dynamical_1996,ford_origins_2008, owen_photoevaporation_2018}. This scenario can bring some Neptunes to large orbital distances, delay their migration, and therefore can protect them from evaporation. Alternatively, high-eccentricity migration can also bring some Neptunes closer to their host stars, and possibly disrupting them through stellar tides. Finally, the planets initially in the Neptune desert can migrate away through tidal and magnetic interactions with their host stars \citep{ahuir_magnetic_2021}. Hence, one way to test these theories requires measuring the planet's stability against photoevaporation through mass-loss rate measurements for a large exoplanet population to derive statistical conclusions.\\  

The ongoing evaporation of exoplanetary atmospheres was first observed through the hydrogen Lyman-$\alpha$ line at UV wavelengths for several hot Jupiters \citep[e.g.,][]{vidal-madjar_extended_2003, lecavelier_des_etangs_temporal_2012, bourrier_atmospheric_2013} and warm Neptunes \citep{ehrenreich_giant_2015,bourrier_hubble_2018,ben-jaffel_signatures_2022}. However, the  interstellar medium (ISM), the geocoronal emission, and the lack of stellar continuum limits the use of the Ly-$\alpha$ line, only observable from space, to measure exoplanet mass-loss rate. \\

The near-infrared helium triplet, predicted earlier on by \cite{seager_theoretical_2000} (see also \citet{oklopcic_new_2018}), has recently been discovered \citep{spake_helium_2018} and is since used to study the upper layers of exoplanet atmosphere from the thermosphere to the exosphere. The exosphere is the outermost atmospheric layer of an exoplanet and is no more gravitationally bound to it. Ground-based near-infrared high-resolution spectrographs (e.g., CARMENES, GIANO, NIRPSEC, or SPIRou) has led to several unambiguous spectrally and temporally resolved detections \citep[e.g.,][]{allart_spectrally_2018, nortmann_ground-based_2018, salz_detection_2018,allart_high-resolution_2019,alonso-floriano_he_2019, kirk_confirmation_2020,guilluy_gaps_2020,spake_posttransit_2021, zhang_escaping_2022,zhang_more_2022,czesla_h_2022}, highlighting the use of the helium triplet as a robust atmospheric tracer. In addition, low resolution observations have confirmed and detected helium signatures \citep[e.g.,][]{spake_helium_2018,mansfield_detection_2018,vissapragada_constraints_2020,paragas_metastable_2021,fu_water_2022}. Most of these detections were obtained for planets orbiting K-dwarfs, which favor the presence of helium particles in their metastable state in exoplanet atmospheres due to their higher extreme-ultraviolet and lower mid-ultraviolet flux \citep{oklopcic_helium_2019}. This is also in agreement with several non-detections for planets orbiting around stars of other spectral types \citep[e.g.,][]{kasper_nondetection_2020,casasayas-barris_carmenes_2021,fossati_gaps_2022}. \\  
In addition to being a powerful atmospheric tracer, the helium triplet is weakly (or not) affected by the ISM absorption \citep{indriolo_interstellar_2009}, the Rossiter-McLaughlin effect, the center-to-limb variation or by stellar activity \citep{cauley_effects_2018}. Therefore, measuring these transitions has yield estimates of the mass-loss rate of tens of exoplanets orbiting K dwarfs. However, the disparateness in the instruments, data reduction pipelines, transmission spectrum extractions, data reproducibility, and modeling framework have prevented homogeneous analyses thus far. Although, homogenous analyses have been led on a handful of exoplanets \citep[e.g.,][]{kasper_nondetection_2020,vissapragada_upper_2022,zhang_detection_2023}, we provide the largest homogeneous analysis of the helium triplet in the atmosphere of eleven exoplanets observed with SPIRou.\\

We describe the instrument and the observations in section\,\ref{sec:obs}, then detail the methods used in section\,\ref{sec:methods}. Section\,\ref{sec:survey} presents the helium analysis for each planet, while section\,\ref{sec:discussion} discuss the general trends that can be drawn from this sample. We conclude in section\,\ref{sec:conclusion}.

\section{Observations}\label{sec:obs}
SPIRou (SPectrom\`etre InfraROUge, \citealt{donati_spirou_2020}) is a fiber-fed near-infrared (0.98$-$2.51\,$\mu$m) echelle spectro-polarimeter installed on the 3.6\,m Canada France Hawaii Telescope in Maunakea (CFHT). It has a high spectral resolution of 70\,000 with 1.9 pixels per resolution element and a pixel sampling of 2.3\,km$\cdot$s$^{-1}$. SPIRou is fed by three fibers: fibers A and B for science with orthogonal polarization and fiber C for reference. SPIRou was already used for atmospheric studies \cite[e.g.][]{boucher_characterizing_2021,pelletier_where_2021}, which reported the presence of non white noise instrumental systematics that might be associated to modal noise for example \citep{oliva_experimental_2019}. Data reduction and analysis are described in section\,\ref{sec:methods}.\\

\begin{figure}
\includegraphics[width=\columnwidth]{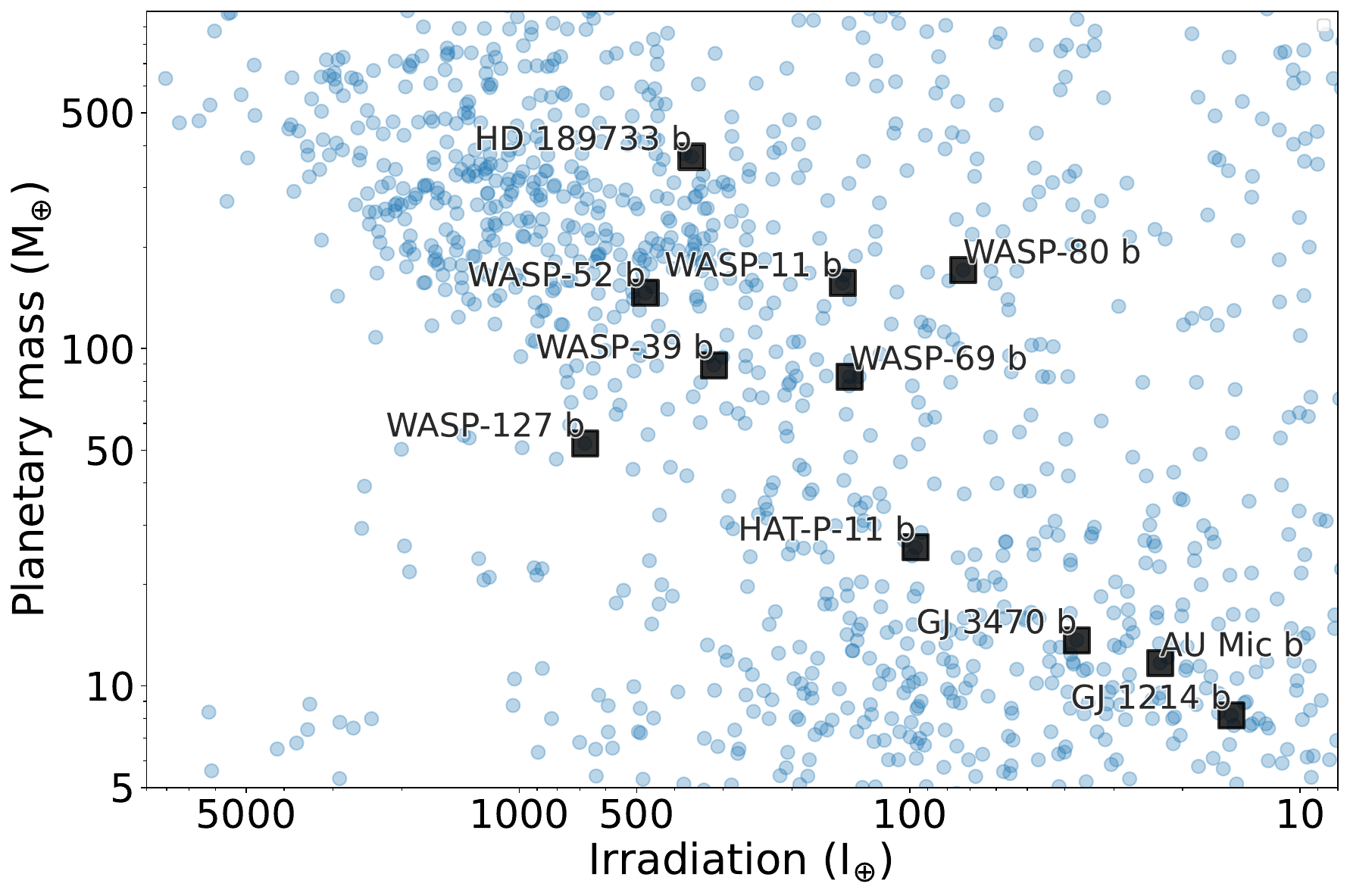}
\caption{Planetary mass - irradiation diagram of the exoplanet population with the studied SPIRou targets highlighted as black squares. The blues dots is the exoplanet population from the NASA exoplanet Archive (December 23rd 2022).}
    \label{fig:Population}
\end{figure}

Transit datasets of 13 exoplanets have been collected with SPIRou as part of the SPIRou Legacy Survey (SLS, PI: Donati) and various programs obtained through Canadian open time or collaborations (namely  AU\,Mic\,b, GJ1214\,b, GJ3470\,b, HAT-P-11\,b, HD1897333\,b, K2-25\,b, TOI-1728\,b WASP-11\,b, WASP-39\,b, WASP-52\,b, WASP-69\,b, WASP-80\,b, and WASP-127\,b). The polarimetry mode was used for the datasets collected on 2019-06-17 for AU\,Mic\,b, 2019-02-18 for GJ\,3470\,b, and 2020-07-03, 2020-07-05, 2020-07-25, and 2021-08-24 for HD\,189733\,b. However, we used the extracted data in the AB mode such that we do not differentiate between polarization (see section\,\ref{sec:apero} and \cite{cook_apero_2022}). We discard K2-25b (19BP40, PI: Donati) due to the very low signal-to-noise ratio (S/N) obtained for each exposure and TOI-1728b (21BC14, PI: Allart) due to a mismatch of the transit window. Figure\,\ref{fig:Population} highlights the targets observed with SPIRou in a planetary mass-irradiation diagram. The sample is mainly composed of hot to warm Jupiters (7) with some warm Neptunes (2) and mini-Neptunes (2) spanning a broad range of stellar ages and orbiting mainly K and M-type stars.  We summarize the observational conditions (i.e. S/N, seeing, and airmass) during planetary transits in table\,\ref{tab:observations}. We note that due to CFHT scheduling constraints for SPIRou and different program strategies, it was not possible to gather more than one transit for some targets, limiting the reproducibility of the results.

\begin{table*}
	\centering
	\caption{Observations summary.}
	\label{tab:observations}
	\begin{tabular}{lccccccccc} 
		\hline
		Target 		& Program ID & PI 		& UT date 	& S/N 				& Seeing & Airmass 	& t$_{\mathrm{exp}}$ [s]	& N$_{\mathrm{exp}}$ & Partial transit\\
		\hline
		AU\,Mic\,b 	& 19AP42		& Donati		& 2019-06-17 & 57 - 91 - 105		& 0.5 - 0.8 - 1.7	& 1.6 - 1.7 - 2.9 		& 122.6 & 116 		& No\\
		GJ1214\,b 	& 20AP40		& Donati		& 2020-05-14 & 8 - 9 - 11 		& ---				& 1.1 - 1.2 - 1.4		& 250.7 & 30 		& No \\
		GJ3470\,b	& 19AP40		& Donati		& 2019-02-18 & 17 - 21 - 25 		& ---				& 1.0 - 1.1 - 1.5		& 434.6 & 32 		& No \\
					& 21BP40		& Donati		& 2021-12-15 & 24 - 38 - 42		& 0.5 - 0.7 - 1.4	& 1.0 - 1.2 - 2.1		& 300.9 & 43 		& No \\
		HAT-P-11\,b	& 21BC09		& Radica		& 2021-08-13 & 98 - 107 - 111	& 0.2 - 0.3 - 0.8	& 1.1 - 1.2 - 1.3		& 323.2 & 44 		& No \\
			        & 21BF19		& Debras		& 2021-08-18 & 69 - 87 - 96		& 0.7 - 1.0 - 1.4 	& 1.1 - 1.2 - 1.6		& 284.2 & 46 		& No\\
		HD189733\,b	& 21BC09		& Donati		& 2018-09-22 & 112 - 120 - 126	& 0.4 - 0.5 - 0.6	& 1.0 - 1.1 - 1.3		& 250.7	& 36			& No\\
			        & 19AP40		& Donati		& 2019-06-15 & 84 - 114 - 125	& 0.4 - 0.7 - 0.9 	& 1.0 - 1.0 - 1.1		& 250.7 & 50 		& No\\
			        & 20AC01		& Deibert	& 2020-07-03 & 12 - 94 - 99		& 0.4 - 0.5 - 0.7 	& 1.1 - 1.2 - 1.4		& 50.1	& 56 		& Yes\\
			     	& 20AC01		& Deibert	& 2020-07-05 & 41 - 63 - 83		& ---				& 1.0 - 1.0 - 1.1		& 94.7	& 52 		& Yes\\
			     	& 20AC01		& Deibert	& 2020-07-25 & 42 - 78 - 92		& ---				& 1.0 - 1.2 - 2.1		& 94.7	& 124 		& No \\
			     	& 21BC16		& Deibert	& 2021-08-24 & 77 - 104 - 118	& 0.5 - 0.7 - 1.0 	& 1.0 - 1.0 - 1.3		& 94.7	& 108 		& No \\
		WASP-11\,b 	& 19BC08		& Boucher	& 2019-10-08 & 15 - 17 - 18 		& ---				& 1.0 - 1.1 - 1.3		& 601.8 & 30 		& No\\
		WASP-39\,b 	& 22AC30		& Pelletier	& 2022-06-04 & 9 - 10 - 12		& 0.4 - 0.7 - 0.9	& 1.1 - 1.1 - 1.0		& 300.9	& 46 		&No\\
					& 22AC30		& Pelletier	& 2022-06-08 & 9 - 12 - 14		& 0.3 - 0.4 - 0.7	& 1.1 - 1.1 - 1.3		& 300.9	& 46 		&No\\
		WASP-52\,b 	& 19BC07		& Boucher	& 2019-10-15 & 7 - 9 - 12 		& ---				& 1.0 - 1.0 - 1.2		& 685.3 & 18 		& No\\
					& 19BC07		& Boucher	& 2019-11-05 & 10 - 13 - 14 		& 0.2 - 0.3 - 0.5 	& 1.0 - 1.0 - 1.3		& 685.3 & 18 		& No\\
		WASP-69\,b 	& 19BP40		& Donati		& 2019-10-13 & 7 - 20 - 26		& 0.5 - 0.7 - 1.4 	& 1.1 - 1.3 - 2.2		& 122.6 & 93 		& No\\
		WASP-80\,b 	& 19BP40 	& Donati		& 2019-10-07 & 10 - 12 - 13		& ---				& 1.1 - 1.1 - 1.7 		& 183.9 & 74 		& No\\
		WASP-127\,b	& 20AP42		& Donati		& 2020-03-11 & 22 - 26 - 28 		& ---				& 1.1 - 1.1 - 1.5 		& 300.9 & 50 		& No\\
					& 21AC02		& Boucher	& 2021-03-22 & 32 - 51 - 56 		& ---				& 1.1 - 1.2 - 1.8 		& 373.3 & 28 		& Yes\\
					& 21AC02		& Boucher	& 2021-05-03 & 14 - 36 - 62 		& 0.3 - 0.4 - 0.6 	& 1.1 - 1.1 - 1.5 		& 501.5 & 28 		& Yes\\
		\hline
	\end{tabular}
\tablefoot{S/N, Seeing, and Airmass values are given as the minimum-median-maximum values of each night. The S/N values are for the echelle order 71, where the helium lines are. Some seeing values are missing due to the lack of measurements. Partial transits are defined as transit with more than a fourth is missing.}
\end{table*}

\section{Methods}\label{sec:methods}

\subsection{Data reduction, spectral extraction and telluric correction}\label{sec:apero}
All data were reduced using A PipelinE to Reduce Observations (\texttt{APERO}; version 0.7.179; \citealt{cook_apero_2022}), the standard SPIRou data reduction software. \texttt{APERO} performs all calibrations and pre-processing to remove detector effects including dark, bad pixel, background, and performs detector non-linearity corrections \citep{artigau_h4rg_2018}, localization of the orders, geometric changes in the image plane, correction of the flat and blaze, hot pixel and cosmic ray correction, wavelength calibration (using both a hollow-cathode UNe lamp and the Fabry-P\'erot \'etalon; \citealt{hobson_spirou_2021}), and removal of diffuse light from the reference fiber leaking into the science channels (when a Fabry-P\'erot is used simultaneously in the reference fiber). This is done using a combination of daily calibrations and reference calibrations. The result is an optimally extracted spectrum of dimensions 4088 pixels (4096 minus 8 reference pixels) with 49 orders, referred to as extracted 2D spectra; \texttt{E2DS}. While the \texttt{E2DS} are produced for the two science fibers (A and B) and the combined flux in the science fibers (AB), we only used the AB extraction as this is the relevant data product for non-polarimetric observations.\\

\texttt{APERO} also provides telluric-corrected versions of the spectra (Artigau et al. in prep) from both the absorption and emission of the Earth's atmosphere. The telluric absorption line correction done in \texttt{APERO} is a two-step process and is briefly outlined here. First, the extracted spectra of both science targets and a large set of rapidly rotating hot stars are fitted with an Earth's transmittance model from TAPAS \citep{bertaux_tapas_2014} that leaves percent-level residuals. Then, from the ensemble of hot star observations, \texttt{APERO} derives a correction model for the residuals with three components for each pixel (optical depths for water, non-water absorbing components, and a constant). This residual model is adjusted to each science observation according to the optical depth of each component from the TAPAS fit. The resulting correction leaves residuals at the level of the PCA-based method of \cite{artigau_telluric-line_2014}, but has the advantage of simplicity and that any spurious point in the data will result in a local error rather than affecting the transmission globally as for a PCA analysis. Finally, a reconstruction of the telluric spectrum is derived using the fitted TAPAS template and the residuals model, for each observed spectrum. The pipeline performs the telluric absorption correction for lines with a transmission down to $\sim$10\% (i.e., with relative depths of 90\% with respect to the continuum), with deeper lines being masked out\footnote{In the spectral order of interest, there are no such deep lines.}. SPIRou does not include a simultaneous sky fiber, so the sky emission correction is done through a high-SNR sky library. A large library of sky spectra has been obtained through the life of the instrument and, from it, a PCA decomposition has been performed. The first 9 components of the sky spectra are fitted to the science data. To avoid subtracting the continuum of stars, this is done through a fit of the derivative of the flux rather than the flux. As sky emission lines are very narrow and have no correspondence in the stellar spectrum, this derivative-fitting is not biased systematically. However, the OH emission lines surrounding the metastable helium triplet are not well modeled and, therefore, not well corrected (residuals up to a few \%) in comparison to the other sky emission lines in the SPIRou spectral range. As described in \cite{oliva_lines_2015} and \cite{czesla_h_2022}, the OH lines around the He triplet are composed of two doublets: 10\,834.3338\,\AA\ ([5-2]Q1e) with 10\,832.412\,\AA\ ([5-2]Q2e) and 10834.241\,\AA\ ([5-2]Q1f) with 10\,832.103\,\AA\ ([5-2]Q2f). The Q1-branch transition lines are the strongest and cannot be distinguished from each other. We, therefore, model these lines as three Gaussians with fixed positions, an amplitude ratio between the Q1 and Q2 lines of 0.0482, a free amplitude and FWHM for the strongest unresolved Q1 lines, and fixed FWHM for the Q2 lines at the element resolution. This model is fitted to the reconstructed sky emission spectra of \texttt{APERO} and then a linear minimization of this best fit with a stellar template is applied for each stellar spectrum. The telluric correction can be seen in Fig.\,\ref{fig:Master_out}.

\subsection{Data analysis}
We follow standard data analysis procedures for studies of exoplanet atmospheres at high resolution as described in detail for example in \cite{wyttenbach_spectrally_2015}, \cite{casasayas-barris_detection_2017}, \cite{allart_spectrally_2018,allart_high-resolution_2019} or \cite{seidel_hot_2020}.\\
Once the stellar spectra are extracted and telluric corrected, we focus our analysis on echelle order 71 (10639-10976\,\AA), where the helium triplet falls at the top of the blaze function. The spectra are moved to the stellar rest frame using the systemic velocity measured by the line-by-line (LBL, \citealt{artigau_line-by-line_2022}) code, normalized using the median flux in two bands (10\,823-10\,826\,\AA\, and 10\,839-10\,842\,\AA), and remaining outliers (e.g., cosmic rays) are sigma replaced following \cite{allart_search_2017}. Spectra obtained before and after transit, hereafter called out-of-transit spectra, are averaged to create a reference master-out spectrum. Figure\,\ref{fig:Master_out} displays the master out for each night of each star before and after applying the telluric correction.

\subsubsection{Transmission spectrum}
To remove stellar features and obtain a transmission spectroscopy map, we divide each spectrum of the time series by the master out. Figure\,\ref{fig:TS_map_night} displays the transmission spectroscopy map for each night of each planet in the stellar rest fame. They are then Doppler-shifted to the planet rest frame based on the parameters in tables\,\ref{tab:star_planet_parameters} and \ref{tab:star_planet_parameters2}. Figure\,\ref{fig:TS_map} shows the transmission spectroscopy map in the planet rest frame for each planet, but averaged over the multiple transits. Partial transits will contribute only to the phases where data were collected. The 1-dimensional transmission spectrum is computed for each night as the average of the transmission spectroscopy map weighted by a modelled white light curve \citep{allart_high-resolution_2019,mounzer_hot_2022}. The \texttt{batman} \citep{kreidberg_batman_2015} package was used to model the white light curve with the parameters from tables\,\ref{tab:star_planet_parameters} and \ref{tab:star_planet_parameters2}, where the quadratic linear limb darkening coefficient have been estimated in the J-band with the \texttt{EXOFAST} \citep{eastman_exofast_2013} code based on the tables of \cite{claret_gravity_2011}. This scaling is necessary to properly take into account the true contribution of the ingress and egress spectra into the transit average transmission spectrum. Finally, and for a given planet, the transmission spectrum of each night are weight averaged by their uncertainties to build the average transmission spectrum. Figure\,\ref{fig:Average_TS} displays the average transmission spectrum around the helium triplet for each planet. 

We neglect the impact of the Rossiter McLaughlin effect \citep{rossiter_detection_1924,mclaughlin_results_1924} and the center-to-limb variation \citep{pierce_solar_1977,neckel_solar_1994} as it was shown to have little impact on the helium lines in several studies \citep[e.g.,][]{allart_spectrally_2018,allart_high-resolution_2019,nortmann_ground-based_2018,salz_detection_2018,fossati_gaps_2022}, which include part of the targets studied here.

\subsubsection{Light curve}

We derive the helium light curve to study the temporal variability of the signal by measuring the excess absorption, assuming a symmetric signal, in a passband of 0.75\,\AA\ centered at 10833.22\,\AA\ for each exposure of the transmission spectroscopy map in the planet rest frame. Figure\,\ref{fig:Average_LC} displays the measured helium light curve for each planet averaged over the multiple transits.

\subsubsection{Detection significance}\label{detect_sign}
The excess absorption is estimated on the transmission spectrum for each transit and the average transmission spectrum by measuring the average signal on a passband of 0.75\,\AA\ centered at 10\,833.22\,\AA. To assess the uncertainty on the measured excess absorption we produced Allan plots (Fig.\,\ref{fig:Allan}) to estimate the contribution of red-noise to the data. We applied the technique described in \cite{winn_transit_2007, cowan_thermal_2012, cubillos_correlated-noise_2017} but instead of having a time-correlated noise source we have a spectrally-correlated noise source. We first estimate the expected Allan curve if our transmission spectrum is solely affected by white noise. We compute the standard deviation of the transmission spectrum excluding the helium triplet (10820-10830 and 10836-10845\,\AA) and scale it for decreasing spectral resolution as $\sqrt{n}$, where $n$ is the number of pixels in the bin. We then binned the transmission spectrum by $n$, compute the root mean square (rms) and repeat the process for a different $n$ size. We then fit the rms in a log-log space to derive the general trend of the noise properties. We then scale our white noise value on the 0.75\,\AA\ bandpass to match the fitted rms. This technique provides a more rigorous estimation of the noise present in the data. We set the detection level as the measured excess absorption divided by the inflated 1-$\sigma$ uncertainty, following the aforementioned noise estimation. In the case of non-detections (below 5-$\sigma$), we report three times the 1-$\sigma$ uncertainty to set the 3-$\sigma$ upper limit. From there, we derive the equivalent opaque radius and the widely used $\delta$R$_p$/H parameter. The latter corresponds to the number of scale heights probed by the equivalent opaque radius. Table\,\ref{tab:results} summarizes these parameters for each planet.

\subsubsection{Bootstrap analysis}
The last test that is performed to confirm the planetary origin of the helium absorption is a bootstrap analysis, also called Empirical Monte Carlo (EMC) simulations \citep[e.g.,][]{redfield_sodium_2008,wyttenbach_spectrally_2015}. It consists of generating three transmission spectrum scenarios (out/out, in/in, and in/out) of 10000 iterations each. The goal is to in produce fake time series' to estimate how likely the measured signal can be built by random noise. For each iteration, the in- and out-of-transit spectra are randomized among the pool of spectra considered in each scenario. Then, the transmission spectrum is built for each of these iterations and the excess absorption is measured as described in \ref{detect_sign}. The in/in and out/out distributions are expected to be centered at zero absorption while the in/out scenario should be close to the excess absorption measured. The results are shown in Fig.\,\ref{fig:Boot_strap}.

\subsection{Modelling}

\subsubsection{Stellar pseudo-signal}\label{model:pseudosignal}
The typical shape of the helium profiles is too broad to spectrally differentiate between a planetary and stellar origin, ; the planet and star signatures overlap for most orbirtal configurations, exception made if the planet is on an eccentric orbit, such as HAT-P-11b \citep{allart_spectrally_2018}. Moreover, it is possible that the planet transits in front of an inhomogeneous stellar surface that can create pseudo signal either in absorption or emission and can partly contribute to the observed He signal \citep[e.g.,][]{rackham_access_2017,salz_detection_2018,guilluy_gaps_2020}. We can consider the stellar disk as two distinct regions with bright and dark stellar patches. The helium absorption line is only produced in the dark region and the planet only transits one of those two regions. If the planet only transits the bright region, the pseudo-absorption signal is maximized. Conversely, if the planet only transits the dark region, the pseudo-emission signal is maximized. To better visualize this effect, we developed a simple toy model that describes these extreme cases and consists of two stellar spectra representing the bright ($F_B$) and dark ($F_D$) regions associated with the fraction of dark regions ($f$, also called filling factor). The observed normalized master out-of-transit spectrum ($\mathrm{F_{out, norm}}$) can be expressed as
\begin{equation}
\mathrm{F_{out, norm}}= (1-f)\cdot F_B +f\cdot F_D .
\end{equation}
$F_B$ and $F_D$ are fitted from 10827 to 10837\,\AA\ to the \ion{Si}{i} line at 10\,830.054\,\AA\ and to the \ion{He}{i} lines of the $\mathrm{F_{out, norm}}$. Following the prescription of \cite{andretta_estimates_2017} and used in \cite{salz_detection_2018}, two superposed Voigt profiles are fitted to the \ion{Si}{i} at the same fixed wavelength, and two Gaussians are fitted to the \ion{He}{i} lines with fixed wavelengths. We fixed the \ion{Si}{i} line profile to be similar between $F_B$ and $F_D$ and we consider that the amplitude of the \ion{He}{i} lines between $F_B$ and $F_D$ are proportional to a constant $\alpha$.  The filling factor is estimated using the relation of \cite{andretta_estimates_2017} (Fig.\,11) with the equivalent width (EW) of the \ion{He}{i} lines at 10\,833\,\AA\. \\
Assuming the planet only transits the bright region, the pseudo-signal can be written as
\begin{equation}
\dfrac{\mathrm{F_{in, norm}}}{\mathrm{F_{out, norm}}}= \dfrac{1}{1-(R_p/R_{\ast})^2}\dfrac{(1-(R_p/R_{\star})^2)\cdot F_B +f\cdot (F_D-F_B)}{(1-f)\cdot F_B +f\cdot F_D},
\end{equation}
which is equivalent to equation 11 of \cite{rackham_access_2017} in the case of ground-based high-resolution normalized spectra.\\
We explored the impact of $\alpha$ and $f$ on the stellar spectrum and the strength of the pseudo-signal for all our targets. The stellar spectra are well reproduced except for low values of $\alpha$ and $f$, i.e. when the stellar helium absorption comes only from a small dark region. The maximum pseudo-signal is produced when all the stellar helium absorption comes from the dark region ($\alpha$=0) independently of the filling factor value selected between $\sim$0.4 and 1.\\

\subsubsection{\texttt{p-winds} modeling}\label{model:pwinds}
The \texttt{p-winds} code \citep{dos_santos_p-winds_2022} is used to calculate the thermospheric structure of the 11 targets and the resulting neutral helium triplet signature. This 1D model is largely based on the formulations of \cite{oklopcic_new_2018} and \cite{lampon_modelling_2020} and we assumed an atmospheric composition of 90\,\% H and 10\,\% He. The density and velocity profiles of the atmosphere are calculated according to the Parker wind approximation assuming an isothermal planetary outflow \citep{parker_dynamics_1958}. To do so, we input the X-EUV spectral energy distribution (over 0-1170\,\AA, \cite{linsky_intrinsic_2014}) of the 11 targets used to calculate the photoionization of H and He, which we calculate in a mutually consistent manner using the formula from \cite{linsky_intrinsic_2014} that predicts the EUV luminosity. This depends on the total X-EUV flux received by the planet which is calculated thanks to the stellar age and the formula from \cite{sanz-forcada_estimation_2011}. The code calculates the density profiles of hydrogen in its neutral and ionized states, and of helium in its neutral, excited, and singly ionized states. The excited helium level corresponds to the metastable transition at 10\,830\,{\r{A}}, which is the signature of interest for which the code calculates theoretical absorption spectra. The absorption signature is compared to the observation in order to estimate the characteristics of the upper atmosphere, such as temperature and mass-loss rate. However, it remains an approximate characterization by considering ideal theoretical spectra calculated at mid transit without taking into account geometrical effects and inhomogeneities of the stellar surface, while it is compared to the observed mean transmission spectra.\\
We explore the input parameter space of the \texttt{p-winds} models for each of the 11 planets, varying the isothermal temperature profile, $T$, and the total atmospheric escape rate, $\dot{m}$, while having a fixed line-of-sight bulk velocity, $v$, and radius value at the top of the model, $r$. The line-of-sight bulk velocity corresponds to an average helium particle motion due to winds in the probe area around the terminator. For the three planets with detected helium lines, we further explore these last two parameters, $v$ and $r$. Previous studies using the p-wind code or similar codes \citep[e.g.,][]{oklopcic_new_2018,lampon_modelling_2020,dos_santos_p-winds_2022,kirk_kecknirspec_2022} seem to not have explored the role of the upper radius boundary used to calculate the thermospheric structure. Yet, this radius is critical for the calculation of the theoretical helium signature. Increasing the radius until the neutral triplet helium density no longer contributes significantly to the absorption signal is rarely consistent with the validity of the model beyond the Roche lobe. Above the Roche lobe, species are no longer gravitationally bound to the planet, which limits the validity of the model. We, therefore, decided to fit the model top radius for targets with neutral triplet helium detection but set an upper limit at the Roche lobe. This arbitrarily limit restricts the amount of neutral triplet helium in the atmosphere and impacts the relative depth between the two neutral triplet helium absorption lines due to the relative radius ratio or the altitude of the optical thickness of the atmosphere. Varying the model top radius cannot be done for non-detection, as it allows finding a model compatible with the data for any temperature and mass-loss rate by reducing the radius to decrease the absorption. We thus set the radius to the Roche lobe to constrain the maximum escape rate for non-detections.\\
We used $\chi^2$ minimization to identify the best-fitting models and their uncertainties. As we found best fits to yield reduced $\chi^2$ larger than unity, likely because of systematic noise in the data, we chose the conservative approach of scaling the error bars of the data by the square root of the reduced $\chi^2$ from the best fit \citep[see, ][]{hebrard_deuterium_2002,lemoine_deuterium_2002}.\\
For planets with detected signals, we provide uncertainties on the best-fit properties at 1-$\sigma$, while for non-detection we provide upper limits at 3-$\sigma$.\\
We limit the parameter space in temperature using the model \cite{salz_simulating_2016} (see also \citet{caldiroli_irradiation-driven_2021, caldiroli_irradiation-driven_2022}) as a function of the gravitational potential of the planet. Below log($-\Phi_G$) = log $GM_{pl}/R_{pl}$ = 13.0 in erg$\cdot$g$^{-1}$ they predict temperatures below 10\,000\,K, while above this limit they predict temperatures below 20\,000\,K. We limit the parameter space in mass-loss using the maximum mass-loss efficiency for a photoionization-driven isothermal Parker wind \citep[e.g.,][]{vissapragada_maximum_2022}.

\section{SPIRou Survey} \label{sec:survey}
Tables\,\ref{tab:star_planet_parameters} and \ref{tab:star_planet_parameters2} summarize the stellar and planetary parameters used for the eleven systems that we observed. In the following subsections, for each exoplanet, we provide a short background history before describing the analysis of the helium triplet and then present our modeling of the transmission spectra.

\begin{table*}
	\centering
	\caption{System parameters of the first eight planets.}
	\label{tab:star_planet_parameters}
	\resizebox{0.95\textwidth}{!}{
	\begin{tabular}{cccccc} 
		\hline
		Parameters 							& Symbol [Unit] 							& AU\,Mic\,b 									& GJ1214\,b 											& GJ3470\,b 											& HAT-P-11\,b \\
		\hline
		\multicolumn{6}{c}{\textit{Star}}\\
		\hline
		Stellar mass 						& M$_{\star}$ [M$_{\odot}$]				& 0.50$\pm$0.03$\color{blue}^{1}$				& 0.178$\pm$0.010$\color{blue}^{3}$	 				& 0.51$\pm$0.06$\color{blue}^{4}$					& 0.802$\pm$0.028$\color{blue}^{6}$	\\
		Stellar radius 						& R$_{\star}$ [R$_{\odot}$]				& 0.86$\pm$0.05$\color{blue}^{18}$				& 0.215$\pm$0.008$\color{blue}^{3}$	 				& 0.48$\pm$0.04$\color{blue}^{4}$					& 0.683$\pm$0.009$\color{blue}^{6}$	\\
		Stellar age 							& [Gyr]									& 0.022$\pm$0.003$\color{blue}^{1}$				& 	3-10$\color{blue}^{14}$							& 1.65$\pm$1.35$\color{blue}^{4}$					& 	6.5$^{+5.9}_{-4.1}$ $\color{blue}^{13}$			\\
		Limb darkening coefficient			& u$_1$ 									& 0.2348											& 0.0775												& 	0.0866											& 0.2673				\\
		Limb darkening coefficient			& u$_2$ 									& 0.3750											& 0.3627												& 	0.3499											& 0.2649				\\
		Stellar metallicity 					& Fe/H [dex]								& 			---									& 0.29$\pm$0.12$\color{blue}^{3}$					& 0.2$\pm$0.1$\color{blue}^{4}$						& 0.31$\pm$0.05$\color{blue}^{6}$	\\		
		Stellar temperature 					& T$_{\mathrm{eff}}$ [K]					& 3700$\pm$100$\color{blue}^{1}$					& 3250$\pm$100$\color{blue}^{3}$						& 3652$\pm$50$\color{blue}^{4}$						& 4780$\pm$50$\color{blue}^{6}$	\\		
		Surface Gravity		 				& log(g) [cgs]							& 4.39$\pm$0.03$\color{blue}^{1}$				& 5.026$\pm$0.040$\color{blue}^{3}$					& 4.658$\pm$0.035$\color{blue}^{4}$					& 4.59$\pm$0.03$\color{blue}^{6}$	\\		
		Spectral type		 				& 										& M1V											& M4V												& M1.5V												& 	K4V						\\		
		\hline
		\multicolumn{6}{c}{\textit{Planet}}\\
		\hline
		Planetary mass 						& M$_{p}$ [M$_{\oplus}$]					& 11.7$\pm$5$\color{blue}^{1}$					& 8.17$\pm$0.43$\color{blue}^{3}$					& 13.9$\pm$1.5$\color{blue}^{4}$						&27.74$\pm$3.11$\color{blue}^{6}$ \\		
		Planetary radius 					& R$_{p}$ [R$_{\oplus}$]					& 3.55$\pm$0.13$\color{blue}^{19}$				& 2.742$\pm$0.17$\color{blue}^{3}$					& 4.57$\pm$0.18$\color{blue}^{4}$					& 4.36$\pm$0.06$\color{blue}^{6}$	\\		
		Density 								& $\rho$ [cgs]							& 1.25$\pm$0.75$\color{blue}^{19}$				& 2.20	$\pm$0.17$\color{blue}^{3}$					& 0.93$\pm$0.56$\color{blue}^{4}$					& 1.84$\pm$0.21								\\		
		Equilibrium temperature 				& T$_{\mathrm{eq}}$ [K]					& 593$\pm$21$\color{blue}^{1}$					& 596$\pm$19$\color{blue}^{3}$						& 615$\pm$16$\color{blue}^{4}$ 						& 787$\pm$11									\\		
		Orbital period 						& P [days]								& 8.463000(2)$\color{blue}^{1}$					& 1.58040433(13)$\color{blue}^{3}$					& 3.3366413(6)$\color{blue}^{4}$						& 4.887802443(34)$\color{blue}^{6}$\\		
		Mid-transit time (-2\,450\,000)		& T$0$ [days]							& 8330.39051(15)$\color{blue}^{1}$				& 5701.413328(66)$\color{blue}^{3}$					& 6677.727712(22)$\color{blue}^{4}$					& 4957.8132067(53)$\color{blue}^{6}$\\		
		Transit duration 					& t$_{14}$ [hours]						& 3.51$\pm$0.1$\color{blue}^{19}$				& 0.8688$\pm$0.0029$\color{blue}^{3}$				& 1.918$\pm$0.024$\color{blue}^{4}$					& 2.3565$\pm$0.0015$\color{blue}^{6}$	\\		
		Semi-amplitude 						& K$_{\star}$ [m$\cdot$s$^{-1}$]			& 5.8$\pm$2.5$\color{blue}^{1}$					& 14.36$\pm$0.53$\color{blue}^{3}$					& 8.21$\pm$0.47$\color{blue}^{4}$					& 12.01$\pm$1.38$\color{blue}^{6}$\\				
		Semi-major axis 						& a [R$_{\star}$]						& 18.95$\pm$0.35$\color{blue}^{19}$				& 14.85$\pm$0.16$\color{blue}^{3}$					& 12.92$\pm$0.72$\color{blue}^{4}$					& 16.50$\pm$0.18$\color{blue}^{6}$\\		
		Inclination 							& $i$ [deg]								& 89.5$\pm$0.3$\color{blue}^{2}$					& 88.7$\pm$0.1$\color{blue}^{3}$						& 88.88$\pm$0.62$\color{blue}^{5}$					& 89.05$\pm$0.15$\color{blue}^{6}$\\		
		Impact parameter 					& b	[R$_{\odot}$]						& 0.17$\pm$0.11$\color{blue}^{19}$				& 0.325$\pm$0.025$\color{blue}^{3}$					& 0.29$\pm$0.14$\color{blue}^{5}$					& 0.209$\pm$0.032$\color{blue}^{15}$\\		
		Eccentricity 						& $e$									& 0												& 0													& 0.114$\pm$0.052$\color{blue}^{4}$					& 0.2644$\pm$0.0006$\color{blue}^{6}$	\\		
		Periastron argument 					& $\omega$								& 90												& 90													& -82.5$\pm$5.7$\color{blue}^{4}$					& 342.186$\pm$0.179$\color{blue}^{6}$	\\		
		\hline
		\hline
		Parameters 							& Symbol [Unit] 							&  WASP-11\,b 									& WASP-39\,b 										& WASP-52\,b 										& WASP-69\,b \\
		\hline
		\multicolumn{6}{c}{\textit{Star}}\\
		\hline
		Stellar mass 						& M$_{\star}$ [M$_{\odot}$]				& 0.81$\pm$0.04$\color{blue}^{7}$				& 0.913$\pm$0.047$\color{blue}^{9}$	 				& 0.87$\pm$0.03$\color{blue}^{10}$					& 0.826$\pm$0.029$\color{blue}^{11}$	\\
		Stellar radius 						& R$_{\star}$ [R$_{\odot}$]				& 0.772$\pm$0.015$\color{blue}^{7}$				& 0.939$\pm$0.022$\color{blue}^{9}$	 				& 0.79$\pm$0.24$\color{blue}^{10}$					& 0.813$\pm$0.028$\color{blue}^{11}$	\\
		Stellar age 							& [Gyr]									& 7.6$^{+6.0}_{-3.5}$ $\color{blue}^{7}$			& 	8.5$^{+4.0}_{-3.4}$ $\color{blue}^{9}$			& 0.4$^{+0.3}_{-0.2}$ $\color{blue}^{10}$				& $\sim$2$\color{blue}^{12}$				\\
		Limb darkening coefficient			& u$_1$ 									& 0.2465											& 0.1767												& 0.2321												& 0.2696							\\
		Limb darkening coefficient			& u$_2$ 									& 0.2721											& 0.2942												& 0.2773												& 0.2617							\\
		Stellar metallicity 					& Fe/H [dex]								& 0.12$\pm$0.09$\color{blue}^{7}$				& 0.01$\pm$0.09$\color{blue}^{9}$					& 0.03$\pm$0.12$\color{blue}^{10}$					& 0.144$\pm$0.077$\color{blue}^{12}$	\\		
		Stellar temperature 					& T$_{\mathrm{eff}}$ [K]					& 4900$\pm$65$\color{blue}^{7}$					& 5485$\pm$50$\color{blue}^{9}$						& 5000$\pm$100$\color{blue}^{10}$					& 4715$\pm$50$\color{blue}^{12}$	\\		
		Surface Gravity		 				& log(g) [cgs]							& 4.569$\pm$0.018$\color{blue}^{7}$				& 4.453$\pm$0.012$\color{blue}^{9}$					& 4.582$\pm$0.014$\color{blue}^{10}$					& 4.535$\pm$0.023$\color{blue}^{12}$	\\		
		Spectral type		 				& 										& K3												&		G8											& K2													& 		K5					\\		
		\hline
		\multicolumn{6}{c}{\textit{Planet}}\\
		\hline
		Planetary mass 						& M$_{p}$ [M$_{\oplus}$]					& 156$\pm$8$\color{blue}^{7}$					& 89.3$\pm$10.2$\color{blue}^{9}$					& 146.2$\pm$6.36$\color{blue}^{10}$					&82.64$\pm$5.88$\color{blue}^{11}$ \\		
		Planetary radius 					& R$_{p}$ [R$_{\oplus}$]					& 11.10$\pm$0.25$\color{blue}^{7}$				& 14.34$\pm$0.45$\color{blue}^{9}$					& 14.24$\pm$0.34$\color{blue}^{10}$					& 11.85$\pm$0.19$\color{blue}^{11}$	\\		
		Density 								& $\rho$ [cgs]							& 0.632$\pm$0.035$\color{blue}^{7}$				& 0.167$\pm$0.023$\color{blue}^{9}$					& 0.29$\pm$0.03$\color{blue}^{10}$					& 0.262$\pm$0.047$\color{blue}^{11}$								\\		
		Equilibrium temperature 				& T$_{\mathrm{eq}}$ [K]					& 992$\pm$14$\color{blue}^{7}$					& 1166$\pm$14$\color{blue}^{9}$						& 1315$\pm$35$\color{blue}^{10}$ 					& 963$\pm$18$\color{blue}^{11}$									\\		
		Orbital period 						& P [days]								& 3.7224790(3)$\color{blue}^{8}$					& 4.0552941(34)$\color{blue}^{9}$					& 1.7497798(12)$\color{blue}^{10}$					& 3.8681390(6)$\color{blue}^{8}$\\		
		Mid-transit time (-2450000)			& T$0$ [days]							& 6200.28683(9)$\color{blue}^{8}$				& 5342.96913(63)$\color{blue}^{9}$					& 5793.68143(9)$\color{blue}^{10}$					& 7176.17789(17)$\color{blue}^{8}$\\		
		Transit duration 					& t$_{14}$ [hours]						& 2.58$\pm$0.91$\color{blue}^{7}$				& 2.80$\pm$0.02$\color{blue}^{9}$					& 1.81$\pm$0.012$\color{blue}^{10}$					& 2.23$\pm$0.03$\color{blue}^{11}$	\\		
		Semi-amplitude 						& K$_{\star}$ [m$\cdot$s$^{-1}$]			& 82.7$\pm$4.2$\color{blue}^{7}$					& 37.9$\pm$5.4$\color{blue}^{9}$						& 109.6$\pm$4.4$\color{blue}^{10}$					& 38.1$\pm$2.4$\color{blue}^{11}$\\				
		Semi-major axis 						& a [R$_{\star}$]						& 12.19$\pm$0.21$\color{blue}^{7}$				& 11.37$\pm$0.24$\color{blue}^{9}$					& 7.38$\pm$0.11$\color{blue}^{10}$					& 12.00$\pm$0.46$\color{blue}^{11}$\\		
		Inclination 							& $i$ [deg]								& 89.03$\pm$0.34$\color{blue}^{7}$				& 87.32$\pm$0.17$\color{blue}^{9}$					& 85.35$\pm$0.2$\color{blue}^{10}$					& 86.71$\pm$0.20$\color{blue}^{11}$\\		
		Impact parameter 					& b	[R$_{\odot}$]						& 0.054$\pm$0.168$\color{blue}^{16}$				& 0.447$\pm$0.055$\color{blue}^{17}$					& 0.60$\pm$0.02$\color{blue}^{10}$					& 0.686$\pm$0.023$\color{blue}^{11}$\\		
		Eccentricity 						& $e$									& 0												& 0													& 0													& 0	\\		
		Periastron argument 					& $\omega$								& 90												& 90													& 90													& 90	\\		
		\hline

	\end{tabular}}
\tablefoot{We used the following references for the parameters: 1: \citet{zicher_one_2022}. 2: \citet{martioli_new_2021}. 3: \citet{cloutier_more_2021}. 4: \citet{kosiarek_bright_2019}. 5: \citet{biddle_warm_2014}. 6: \citet{allart_spectrally_2018}. 7: \citet{mancini_gaps_2015}. 8: \citet{kokori_exoclock_2022}. 9: \citet{mancini_gaps_2018}. 10: \citet{hebrard_wasp-52b_2013}. 11: \citet{casasayas-barris_detection_2017}. 12: \citet{anderson_three_2014}. 13: \citet{bakos_hat-p-11b_2010}. 14: \citet{charbonneau_super-earth_2009}. 15: \citet{huber_discovery_2017}. 16: \citet{west_sub-jupiter_2009}. 17: \citet{maciejewski_new_2016}. 18: \citet{gallenne_probing_2022}. 19: \citet{szabo_transit_2022}.}
\end{table*}
\begin{table*}
	\centering
	\caption{Continuation of table\,\ref{tab:star_planet_parameters} for the three remaining targets.}
	\label{tab:star_planet_parameters2}
	\begin{tabular}{cccccc} 

		\hline
		Parameters 							& Symbol [Unit] 							&WASP-80\,b										& WASP-127b 											& HD189733\,b 															&   \\
		\hline
		\multicolumn{6}{c}{\textit{Star}}\\
		\hline
		Stellar mass 						& M$_{\star}$ [M$_{\odot}$]				&0.577$\pm$0.054$\color{blue}^{2}$	 			&0.960$\pm$0.023$\color{blue}^{3}$ 					& 0.806$\pm$0.048$\color{blue}^{5}$ 							&  \\
		Stellar radius 						& R$_{\star}$ [R$_{\odot}$]				&0.586$\pm$0.018$\color{blue}^{2}$	 			&1.303$\pm$0.037$\color{blue}^{3}$ 					& 0.756$\pm$0.018$\color{blue}^{5}$ 							&  \\
		Stellar age 							& [Gyr]									&<0.2$\color{blue}^{2}$							&9.656$\pm$1.002$\color{blue}^{3}$ 					& 6.8$\pm$5.2$\color{blue}^{5}$ 								&  \\
		Limb darkening coefficient			& u$_1$ 									&0.2146											&0.1365 												&0.2248  																		&  \\
		Limb darkening coefficient			& u$_2$ 									&0.2830 											&0.2990 												&0.2795  																		&  \\
		Stellar metallicity 					& Fe/H [dex]								&-0.13$\pm$0.17$\color{blue}^{2}$				&-0.19$\pm$0.01$\color{blue}^{3}$ 					& -0.03$\pm$0.08$\color{blue}^{5}$ 							&  \\		
		Stellar temperature 					& T$_{\mathrm{eff}}$ [K]					&4143$\pm$94$\color{blue}^{2}$					&5842$\pm$13$\color{blue}^{3}$						& 5040$\pm$50$\color{blue}^{5}$ 								&  \\		
		Surface Gravity		 				& log(g) [cgs]							&4.663$\pm$0.016$\color{blue}^{2}$				&4.23$\pm$0.02$\color{blue}^{3}$ 					& 4.587$\pm$0.015$\color{blue}^{5}$ 							&  \\		
		Spectral type		 				& 										&K7V												&G5  												& K2V  																			&  \\		
		\hline
		\multicolumn{6}{c}{\textit{Planet}}\\
		\hline
		Planetary mass 						& M$_{p}$ [M$_{\oplus}$]					&171$\pm$11$\color{blue}^{2}$					&52.35$\pm$6.8$\color{blue}^{4}$						&363.582$\pm$18.116$\color{blue}^{5}$  					&  \\		
		Planetary radius 					& R$_{p}$ [R$_{\oplus}$]					&11.2$\pm$0.335$\color{blue}^{2}$				&14.69$\pm$0.33$\color{blue}^{4}$ 					&12.756$\pm$0.303$\color{blue}^{5}$  						&  \\		
		Density 								& $\rho$ [cgs]							&0.717$\pm$0.039$\color{blue}^{2}$				&0.097$\pm$0.013$\color{blue}^{4}$ 					&0.963$\pm$0.088$\color{blue}^{5}$  							&  \\		
		Equilibrium temperature 				& T$_{\mathrm{eq}}$ [K]					&825$\pm$19$\color{blue}^{2}$					&1400$\pm$24$\color{blue}^{3}$ 						& 1201$\pm$13$\color{blue}^{5}$ 								&  \\		
		Orbital period 						& P [days]								&3.06785271(19)$\color{blue}^{1}$				&4.17806203(88)$\color{blue}^{4}$					&2.21857519(14) $\color{blue}^{1}$  		&  \\		
		Mid-transit time (-2450000)			& T$0$ [days]							&6671.49615(4) $\color{blue}^{1}$				&6776.62124(28)$\color{blue}^{4}$ 					&4403.677711(25)  $\color{blue}^{1}$	&  \\		
		Transit duration 					& t$_{14}$ [hours]						&2.131$\pm$0.003$\color{blue}^{2}$				&4.353$\pm$0.014$\color{blue}^{4}$ 					& 1.8036$\pm$0.0023$\color{blue}^{6}$  					&  \\		
		Semi-amplitude 						& K$_{\star}$ [m$\cdot$s$^{-1}$]			&109.0$\pm$4.4$\color{blue}^{2}$					&22$\pm$3$\color{blue}^{3}$ 							&205$\pm$6$\color{blue}^{5}$   									&  \\		
		Semi-major axis 						& a [R$_{\star}$]						&12.63$\pm$0.13$\color{blue}^{2}$				&7.81$\pm$0.11$\color{blue}^{4}$ 					&8.81$\pm$0.06$\color{blue}^{5}$  								&  \\		
		Inclination 							& $i$ [deg]								&89.02$\pm$0.1$\color{blue}^{2}$					&87.84$\pm$0.36$\color{blue}^{4}$ 					&85.58$\pm$0.06$\color{blue}^{5}$   							&  \\		
		Impact parameter 					& b	[R$_{\odot}$]						&0.215$\pm$0.022$\color{blue}^{2}$				&0.29$\pm$0.045$\color{blue}^{4}$ 					&0.680$\pm$0.005$\color{blue}^{5}$   							&  \\		
		Eccentricity 						& $e$									&0												&0 													&0  																				&  \\		
		Periastron argument 					& $\omega$								&90												&90 													&90  																				&  \\		
		\hline
	\end{tabular}
\tablefoot{We used the following references for the parameters: 1: \citet{kokori_exoclock_2022}. 2: \citet{triaud_wasp-80b_2015}. 3: \citet{allart_wasp-127b_2020}. 4: \citet{seidel_hot_2020}. 5: \citet{torres_improved_2008}. 6: \cite{baluev_benchmarking_2015}.}
\end{table*}

\begin{figure*}
\includegraphics[width=\textwidth]{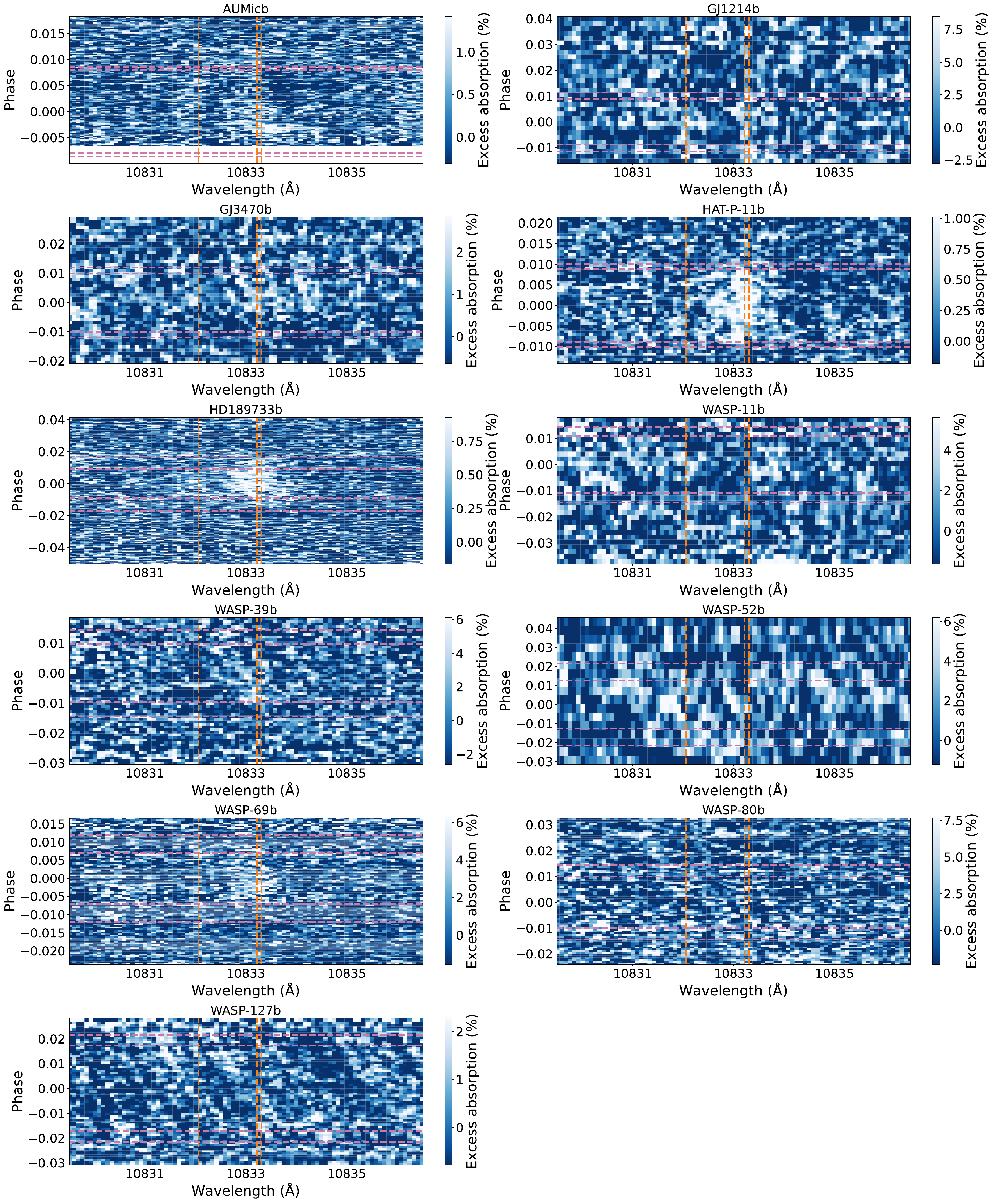}
\caption{Transmission spectroscopy maps in the planet rest frame for the eleven planets. Detections are visible for HAT-P-11\,b, HD\,189733\,b and WASP-69\,b as white clusters. The pink dashed horizontal lines are the contact lines from bottom to top: t$_1$, t$_2$, t$_3$, and t$_4$. The vertical orange dashed lines indicate the helium lines' positions.}
    \label{fig:TS_map}
\end{figure*}

\begin{figure*}
\includegraphics[width=\textwidth]{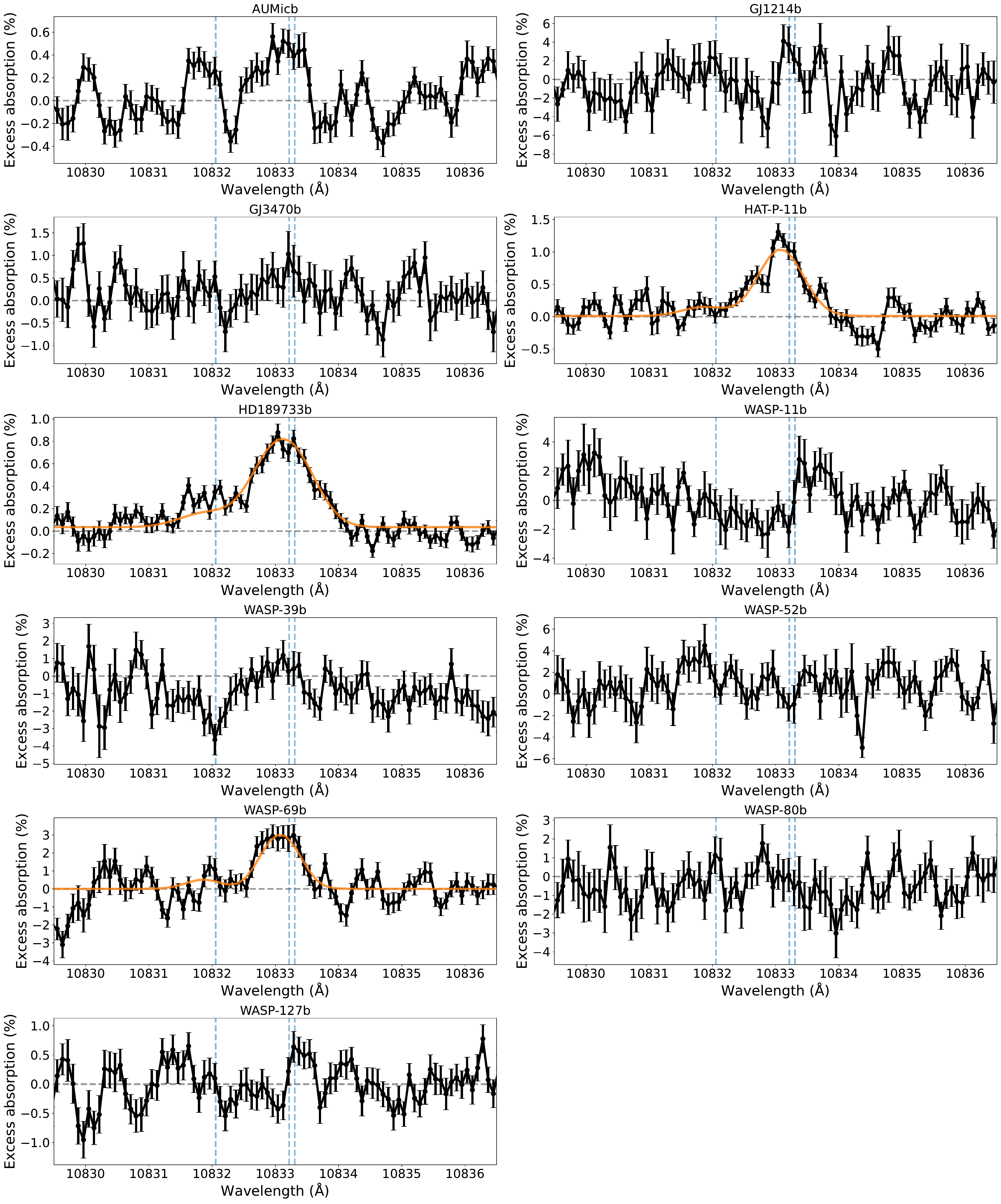}
\caption{Average transmission spectra in the planet rest frame for the eleven planets. The vertical blue dashed lines indicate the helium lines' positions. The horizontal grey dashed line represents null excess absorption. The orange curves are the best-fit model from p-winds obtained for the three detections: HAT-P-11\,b, HD\,189733\,b, and WASP-69\,b.}
    \label{fig:Average_TS}
\end{figure*}

\begin{figure*}[h]
\centering
\includegraphics[width=0.95\textwidth]{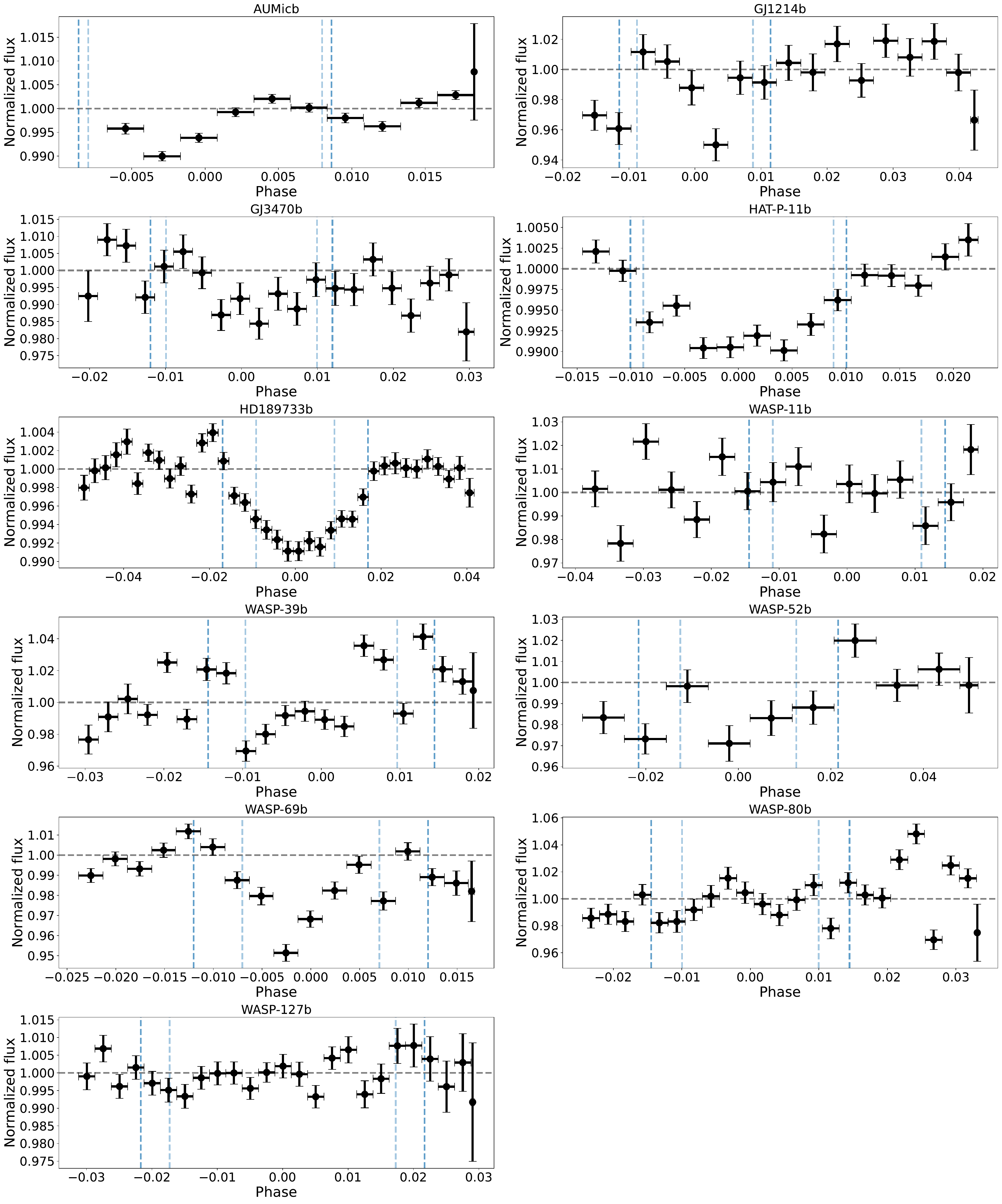}
\caption{Average excess helium light curves for the eleven planets. The blue dashed vertical lines are the contact lines from left to right: t$_1$, t$_2$, t$_3$, and t$_4$. The grey horizontal dashed line is the continuum level.}
    \label{fig:Average_LC}
\end{figure*}

\subsection{AU\,Mic\,b}
\subsubsection{Background}
 AU\,Mic\,b is the inner planet of a two Neptune-sized system orbiting a young M dwarf discovered with TESS and monitored by several radial velocity (RV) spectrographs \citep{plavchan_planet_2020,klein_investigating_2021,martioli_new_2021,zicher_one_2022}. With an age of 22$\pm$3\,Myr, this system still hosts an edge-on debris disc \citep{kalas_discovery_2004,boccaletti_fast-moving_2015,vizgan_multiwavelength_2022} and has an intense magnetic activity cycle. It was shown \citep[e.g.,][]{martioli_spin-orbit_2020,palle_he_2020,addison_youngest_2021} that the planet b has an aligned orbit, and thus, might have formed and migrated within the disc. Therefore, it is thought that  AU\,Mic\,b, and c are progenitors of the super-Earth/mini-Neptune population and key targets for in-depth characterization. Such a young planetary system is of particular importance to understand how planets and their atmospheres evolve. No detections of atomic or molecular species have been reported in the literature, but attempts in the visible have been made with ESPRESSO \citep{palle_he_2020}. In addition, \cite{hirano_limits_2020} used IRD and NIRSPEC data to study the presence of metastable helium and set an upper limit on the equivalent width of 3.7\,m\AA\ at 99\,\%. The lack of atmospheric detections observed could be linked to the stellar wind confining the planet's atmosphere outflow \cite{carolan_dichotomy_2020}.
\subsubsection{Helium triplet}
The only transit of  AU\,Mic\,b observed with SPIRou has no baseline before transit and is missing data until after ingress, due to high airmass during constraints. The telluric lines are redshifted from the helium triplet (Fig.\,\ref{fig:Master_out}). A variable excess absorption feature is visible at the position of the helium triplet, but the width and intensity evolve across the transit with a maximum of absorption before mid-transit (Fig.\,\ref{fig:TS_map}, \ref{fig:Average_TS} and \ref{fig:Average_LC}). The measured excess absorption on the transmission spectrum is 0.37\,$\pm$\,0.09\,\% (4.3\,$\sigma$) assuming the noise properties derived from the Allan plot (Fig.\,\ref{fig:Allan}). Indeed, similar structures are visible at different wavelengths (Fig.\,\ref{fig:Average_TS}). These structures are not caused by telluric contamination (see Fig.\,\ref{fig:Master_out}). However, it can be expected that young active stars have variable stellar features and it is, therefore, not possible to claim any robust detection of helium for  AU\,Mic\,b with only one transit. We set the 3-$\sigma$ upper limit on the presence of helium at a conservative $<$0.26\,\% following the procedure of section\,\ref{detect_sign}, which is in agreement with \cite{hirano_limits_2020}. It is also possible that due to the high stellar activity of  AU\,Mic, the master out spectrum is not representative enough of the stellar features over the transit duration. We, therefore, call for more observations of the system to confirm the signature and average out the stellar activity.

\subsection{GJ\,1214\,b}
\subsubsection{Background}
GJ\,1214\,b is a warm mini-Neptune orbiting a nearby M dwarf \citep{charbonneau_super-earth_2009}. Its density is in good agreement with a water-rich composition and a hydrogen-helium envelope, which encouraged in-depth analysis of its atmosphere. However, \cite{kreidberg_clouds_2014} revealed a featureless near-infrared spectrum obtained with the Hubble Space Telescope (HST), even with exquisite precision. The authors ruled out numerous compositions and concluded that the lower atmosphere of GJ1214\,b is dominated by clouds. Nonetheless, recent attempts have been performed to detect the thermosphere and exosphere of the planet (layers well above the cloud deck) through the helium triplet. \cite{crossfield_upper_2019}, \cite{petit_dit_de_la_roche_upper_2020} and \cite{kasper_nondetection_2020} reported only upper limits on the presence of He, while \cite{orell-miquel_tentative_2022} reported a tentative detection at 4.6$\sigma$. It is interesting to compare the two last results as they have been obtained at high resolution with Keck/NIRSPEC \citep{kasper_nondetection_2020} and CARMENES \citep{orell-miquel_tentative_2022}. The upper limit set with NIRSPEC is $\sim$0.13\,$\%$ at the 90\,$\%$ confidence interval obtained for one transit, while the detection obtained with CARMENES is of 2.1\,$\pm$\,0.5\,$\%$ and was also obtained for one transit. \cite{orell-miquel_tentative_2022} proposed that the discrepancy might be caused by the telluric contamination of the nearby OH and H$_2$O lines and their poor correction. The authors scheduled their CARMENES transits to avoid such contamination, and showed that the H$_2$O line is superposed to the helium triplet in the NIRSPEC data. However, \cite{spake_non-detection_2022} reported an upper limit of $\sim$1.22\,$\%$ at the 95\,$\%$ confidence interval by observing one transit of GJ\,1214\,b with NIRSPEC at a time of the year where there is no telluric contamination. Their results are in clear contradiction with \cite{orell-miquel_tentative_2022} and could be explained by instrumental or reduction systematics \citep{radica_revisiting_2022}, or by strong variability from the star or in the planet's atmosphere.

\subsubsection{Helium triplet}
The time series of GJ\,1214\,b observed with SPIRou spans the full transit with a baseline before and after. The telluric lines are shifted to the red from the helium triplet and do not overlap with the planetary track (Fig.\,\ref{fig:Master_out}). The transmission spectrum (Fig.\,\ref{fig:Average_TS}) is impacted by some systematics and the helium light curve (Fig.\,\ref{fig:Average_LC}) has some variability before and during transit, but the bootstrap analysis (Fig.\,\ref{fig:Boot_strap}) reveals similar distribution with no significant excess absorption for any of the three scenarios. The measured excess absorption on the transmission spectrum is of 1.59\,$\pm$\,0.97\,\%, and the 3-$\sigma$ upper limit on the presence of helium of $<$2.92\,\%, which is not constraining enough to settle the difference between the detection of \cite{orell-miquel_tentative_2022} and the non-detections of \cite{kasper_nondetection_2020} and \cite{spake_non-detection_2022}. Despite having similar instrument and telescope, our results are less sensitive than \cite{orell-miquel_tentative_2022} but are due to the lower S/N of the SPIRou data.

\subsection{GJ3470\,b}

\subsubsection{Background}
GJ3470\,b is a warm Neptune orbiting a nearby M dwarf \citep{bonfils_hot_2012} on an eccentric polar orbit \citep{stefansson_warm_2022}. \cite{benneke_sub-neptune_2019} revealed a low-metallicity, hydrogen-dominated atmosphere with the detection of water, but depleted in methane. One possibility proposed by the authors is the presence of an unknown planet that could have caused tidal heating and pushed the atmosphere to be CO-dominated. The eccentric polar orbit of GJ3470\,b could be an additional consequence of an unknown companion at long period \citep{stefansson_warm_2022}. In addition, \cite{bourrier_hubble_2018} revealed through the detection of neutral hydrogen that the upper atmosphere extends beyond the Roche lobe, is elongated in the direction of the planet's motion, and strongly escapes into space. This could indicate that GJ3470\,b may have lost 4 to 35\,\% of its current mass over its lifetime ($\sim$2Gyr). Metastable helium has also been detected in the upper atmosphere of this planet \citep{ninan_evidence_2020,palle_he_2020}. The latter reported detection of 1.5$\pm$0.3\,\% maximum excess 	absorption with CARMENES and derived a mass-loss rate of the same magnitude as \cite{bourrier_hubble_2018}.

\subsubsection{Helium triplet}
The two time-series observed with SPIRou span the full transit with baselines before and after transit. Telluric lines of OH and water overlap with the helium triplet for the first night but are redshifted for the second night (Fig.\,\ref{fig:Master_out}). In addition, the second time series was observed under better weather conditions. The transmission spectra and helium light curves of both nights are in very good agreement with each other. From the transmission spectroscopic map (Fig.\,\ref{fig:TS_map}), we can see an overall increase of excess absorption on a broad wavelength range from after ingress until the end of the time series, independently of the transit. This effect is visible in the helium light curve (Fig.\,\ref{fig:Average_LC}). However, the averaged transmission spectrum (Fig.\,\ref{fig:Average_TS}) does not have significant excess absorption, which is in agreement with the bootstrap analysis (Fig.\,\ref{fig:Boot_strap}) of the two time-series. The measured excess absorption on the transmission spectrum is 0.55\,$\pm$\,0.21\,\% (2.6$\sigma$). We put the 3-$\sigma$ upper limit on the presence of helium at $<$0.63\,\% as it is difficult to differentiate the observed broad feature between a planetary origin or noise structure. Our result is in disagreement with the detections reported by \cite{ninan_evidence_2020} and \cite{palle_he_2020}, even once they are integrated over the same 0.75\,\AA\ bandpass  ($\sim$1.2\,\% for \cite{palle_he_2020}). We also performed an injection-recovery test by adding to our transmission spectrum a Gaussian of amplitude 1.5\,\% and FWHM of 1\,\AA\ following the result of \cite{palle_he_2020}. The measured excess absorption on this injected data is 1.62\,$\pm$\,0.21\% (7.7$\sigma$), which confirm the tension with  the literature. More data are needed to mitigate non-white noise source and confirm or infirm the presence of metastable helium in the atmosphere of GJ3470\,b.

\subsection{HAT-P-11\,b}

\subsubsection{Background}
HAT-P-11\,b is the inner planet, a warm Neptune \citep{bakos_hat-p-11b_2010}, in a two-planet system \citep{yee_hat-p-11_2018} around a K dwarf star on an eccentric misaligned orbit \citep{winn_oblique_2010}, with properties similar to GJ3470\,b. \cite{fraine_water_2014} and \cite{chachan_hubble_2019} reported the detection of water and methane in its lower atmosphere with high-altitude clouds and a low metallicity, which is in contradiction to the metallicity-mass trend known for the Solar system planets. It is even more striking that the star has a super-solar metallicity. A possible scenario is that metals stopped being accreted before the envelope formed \citep{thorngren_bayesian_2018}. This was further supported by \cite{ben-jaffel_signatures_2022} who reported a low metallicity atmosphere through a panchromatic UV approach. In addition, the authors measured the escape of neutral hydrogen and the presence of a cometary-like tail. \cite{allart_spectrally_2018} and \cite{mansfield_detection_2018} detected the presence of metastable helium at near-infrared wavelengths with CARMENES and HST. Due to the high resolution of CARMENES, \cite{allart_spectrally_2018} resolved the helium lines and measured an excess absorption of 1.08\,$\pm$\,0.05\,\%  on a 0.75\,\AA\ passband with some variability between their two transits (0.82\,$\pm$\,0.09\,\% and 1.21\,$\pm$\,0.06\,\%). They also constrained the presence of helium to the thermosphere at high temperatures (or low mean molecular weight) with the presence of strong day-to-night side winds and without a strong mass-loss rate.
\subsubsection{Helium triplet}
The two transits of HAT-P-11\,b are well observed with a baseline before and after transit. Due to the high systemic velocity of the system, there is no overlap with telluric lines (Fig.\,\ref{fig:Master_out}). A clear repeatable signature is visible during the transit and is slightly blue-shifted from the expected position of the helium triplet in the planetary rest frame, which cannot be mistaken with the stellar rest frame due to the planet eccentricity (Fig.\,\ref{fig:TS_map}, \ref{fig:Average_TS} and \ref{fig:Average_LC}). The helium light curve does not significantly extend beyond the transit duration in agreement with \cite{allart_spectrally_2018}. In addition, the two transits show similar helium line shapes and light curve with no significant temporal variation. The measured excess absorption on the transmission spectrum is 0.76\,$\pm$\,0.07\,\% (11\,$\sigma$) with a maximum of excess absorption at $\sim$1.3\,\%.  The excess absorption is significantly below the reported average excess absorption measured by \cite{allart_spectrally_2018}, but in agreement with the value reported for their first transit of 0.82\,$\pm$\,0.09\,\%. 

\subsection{HD189733\,b}

\subsubsection{Background}
HD189733\,b is a hot Jupiter orbiting a relatively active K dwarf \citep{bouchy_elodie_2005}. Due to its host star brightness, it is one of the most studied exoplanets from its lower atmosphere to its exosphere. Multiple detections of molecules (such as H$_2$O and CO) have been reported both at low- and high-resolution \citep[e.g.,][]{birkby_detection_2013,de_kok_detection_2013,mccullough_water_2014,sing_continuum_2016,brogi_rotation_2016,brogi_exoplanet_2018,alonso-floriano_multiple_2019,cabot_robustness_2019,boucher_characterizing_2021}. Detection of atomic species probing the higher atmospheric layers have also been reported, including Na \citep[e.g.,][]{redfield_sodium_2008,wyttenbach_spectrally_2015}, K \citep[e.g.,][]{keles_potassium_2019}, H (through H-$\alpha$, \citep[e.g.,][]{cauley_optical_2015,cauley_variation_2016,cauley_evidence_2017} and Lyman-$\alpha$, \citep{lecavelier_des_etangs_evaporation_2010, lecavelier_des_etangs_temporal_2012, bourrier_atmospheric_2013} or He \citep{salz_detection_2018,guilluy_gaps_2020,zhang_more_2022}. The helium triplet has been observed from 2016 to 2020 with three different high-resolution spectrographs (CARMENES, GIANO, and Keck/NIRSPEC) for a total of 9 transits. The three studies all report a compact metastable helium atmosphere probing similar atmospheric layers ($\sim$1.2\,R$_P$) and dynamic (blueshift of $\sim$3-4\,km$\cdot$s$^{-1}$) than the Sodium doublet. However, it was put in evidence in \cite{zhang_more_2022} that the excess absorption varies between epochs and instruments: 0.617$\pm$0.017\,\% for CARMENES \citep{salz_detection_2018}, 0.508$\pm$0.015\,\% for GIANO \citep{guilluy_gaps_2020} and  0.420$\pm$0.013\,\% for NIRPSEC \citep{zhang_more_2022}. These variations are unlikely due to starspot occultation but could be caused by instrumental systematics, unocculted stellar active regions, the planet's atmospheric outflow, shear instability, or stellar flares increasing the star's XUV flux \citep{wang_metastable_2021,wang_metastable_2021-1,hazra_impact_2022}.
\subsubsection{Helium triplet}
\begin{figure}
\includegraphics[width=\columnwidth]{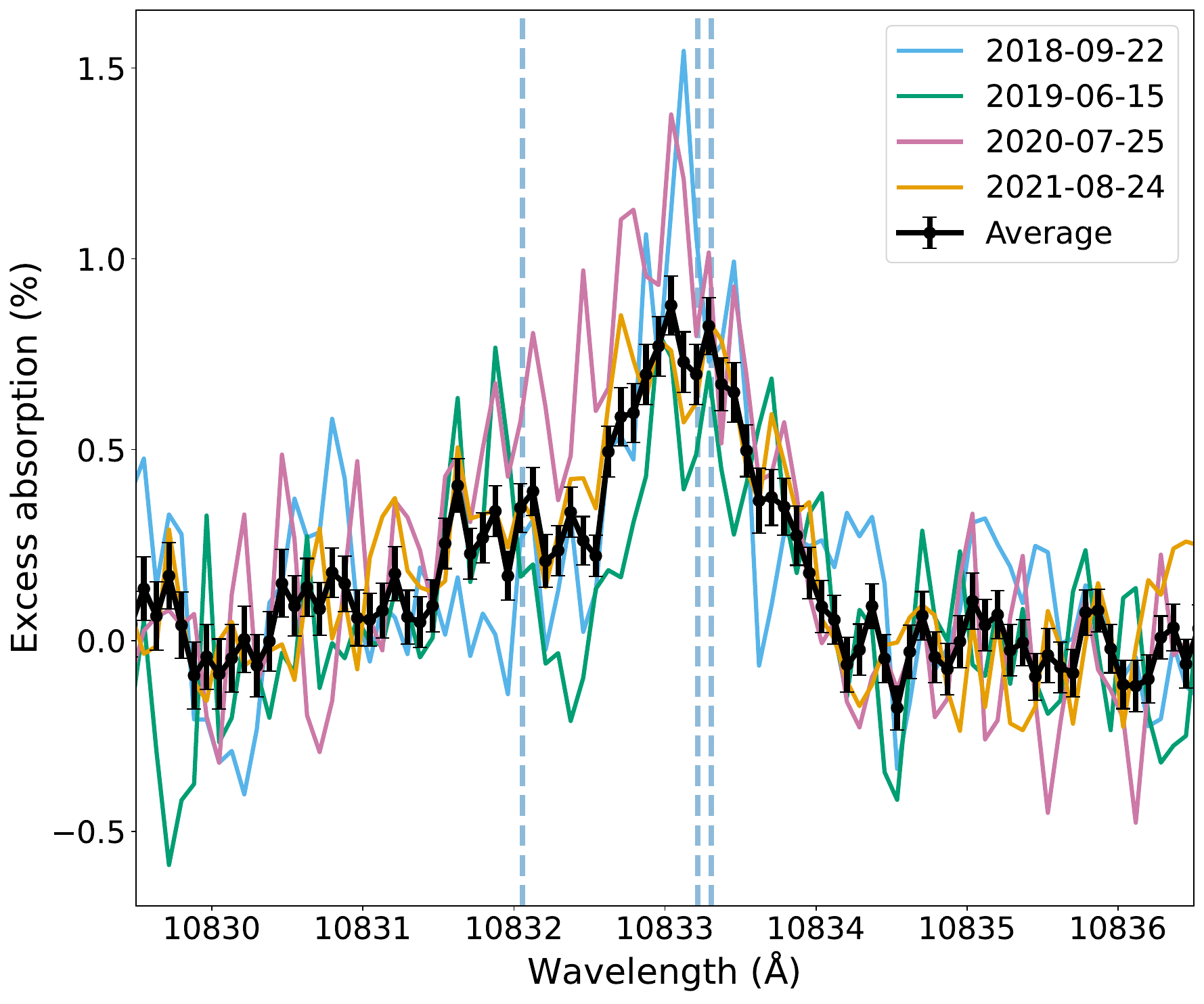}
\caption{Transmission spectra of HD\,189733\,b for the nights where the full transit was observed, respectively 2018-09-22 (blue), 2019-06-15 (green), 2020-07-25 (pink), and 2021-08-24 (orange).}
    \label{fig:TS_HD189733}
\end{figure}

A total of six transit time series' of HD\,189733\, were observed with SPIRou from 2018 to 2021. Two of them observed on 2020-07-13 (night 3) and 2020-07-05 (night 4) are partial transits with respectively only egress and before mid-transit spectra. The transit of 2021-08-24 (night 6) has no after-transit baseline. The remaining transits are well covered with before and after baseline. The telluric contamination is negligible for all nights as either the strong OH component is very shallow or is far away from the helium line. A clear excess absorption feature is detected (Fig.\,\ref{fig:TS_map}, \ref{fig:Average_TS} and \ref{fig:Average_LC}) during the transit of HD\,189733\,b at the expected position of the helium lines, but cannot be disentangled between the stellar and planetary rest frame. The signature is slightly blue-shifted in the planetary rest frame and the two components of the helium doublet are visible with a contrast ratio of $\sim$2. Despite some large variability in the helium light curve before the transit, the excess absorption is well contained to the transit duration. The measured excess absorption on the transmission spectrum is 0.69$\pm$0.04\,\% (17\,$\sigma$) with a maximum at $\sim$0.9\,\%. We report in table\,\ref{tab:results_HD189} the excess absorption measured for each night for the bandpass of 0.75\,\AA\, but also for a 40\,km$\cdot$s$^{-1}$ (1.44\,\AA) bandpass to allow comparison with the previous results of \cite{zhang_more_2022}. The measured excess absorption over this bandpass on the average transmission spectrum differs from the results of NIRSPEC at 3-$\sigma$, GIANO at 0.1-$\sigma$, and CARMENES at 3.6-$\sigma$. The variability in the signal strength is not due to reduction artifact, or Earth's atmosphere residuals as significant variations are measured for different transits obtained with the same instrument (GIANO and SPIRou) and the same data reduction. To further explore this variation in the signal strength, we compare in Fig.\,\ref{fig:TS_HD189733} the transmission spectra obtained for the nights where the complete transit was observed (2018-09-22, 2019-06-15, 2020-07-25 and 2021-08-24). We note that for the transit of 2018-09-22 (blue), the weak component of the helium triplet has no excess absorption while the strong component has more excess absorption than the other nights. The transmission spectrum of 2019-06-15 (green) has less excess absorption in the main component and a clear lack of absorption between the two components. The transmission spectrum of 2020-07-25 (pink) has larger noise structures and there are no clear distinctions between the two components of the helium triplet. These variations of the helium line shape can have different origins such as instrumental systematics, the optical thickness of the outflow or the presence of strong blueshifted helium gas. It is beyond the scope of this paper to investigate the causes of these variations.

\begin{table}
	\centering
	\caption{Summary of the helium triplet excess absorption for each transit of HD\,189733\,b.}
	\label{tab:results_HD189}
	\begin{tabular}{cccc} 

		\hline
				 		&  1-pixel  			&\multicolumn{2}{c}{Excess Absorption [\%], bandpass:}													\\
		Night 			& dispersion [\%] 	& 0.75\,\AA\ 		& 1.44\,\AA\ (40\,km$\cdot$s$^{-1}$)  	\\
		\hline
		\hline
		2018-09-22		& 0.23 				& 0.85$\pm$0.10 	 	& 0.45$\pm$0.08		 											\\
		2019-06-15 	 	& 0.19 				& 0.52$\pm$0.09 	 	& 0.37$\pm$0.07													\\	
		2020-07-03 	 	& 0.28 				& 0.61$\pm$0.14	 	& 0.29$\pm$0.10													\\
		2020-07-05	 	& 0.32 				& 0.58$\pm$0.15 		& 0.57$\pm$0.11 	 												\\
		2020-07-25	 	& 0.22				& 0.90$\pm$0.07 		& 0.75$\pm$0.04 	 												\\
		2021-08-24 	 	& 0.15 				& 0.68$\pm$0.07 	 	& 0.61$\pm$0.05													\\
		Average 		 	& 0.09				& 0.69$\pm$0.04 	 	& 0.51$\pm$0.03	 												\\
		\hline
	\end{tabular}
\end{table}

\subsection{WASP-11\,b}

\subsubsection{Background}
WASP-11b, also known as HAT-P-10b, is a hot Jupiter orbiting an inactive K dwarf \citep{west_sub-jupiter_2009,bakos_hat-p-10b_2009}. It has an aligned orbit, as is the case for many hot Jupiters \citep{albrecht_stellar_2022}. No studies have been reported on its atmosphere.
\subsubsection{Helium triplet}
The time series of WASP-11\,b observed with SPIRou spans the full transit with a baseline before and after. We removed the two first exposures due to high variability in the stellar spectrum. The telluric lines are redshifted relative to the helium triplet (Fig.\,\ref{fig:Master_out}). The transmission spectrum (Fig.\,\ref{fig:Average_TS}) exhibits a slightly decreasing slope from 10\,830 to 10\,833\,\AA\, which is likely due to the \ion{Si}{i} stellar line at 10830\,\AA. It also has some excess absorption features around 10\,833.7\AA\ (redward of the helium triplet), which seems to be associated with a few of exposures before mid-transit (Fig.\,\ref{fig:TS_map}). However, the helium light curve (Fig.\,\ref{fig:Average_LC}) is stable across the time series and the bootstrap analysis (Fig.\,\ref{fig:Boot_strap}) reveals similar distribution with no excess absorption for the out-out, in-in, and in-out scenarios. The measured excess absorption on the transmission spectrum is -0.09\,$\pm$\,0.52\,\%, consistent with no absorption. The 3-$\sigma$ upper limit on the presence of helium is set at $<$1.56\,\% due to the observed systematics. At the difference of the our other datasets, the noise structures disappear and tend toward white noise for larger bin (Fig.\,\ref{fig:Allan}). 

\subsection{WASP-39\,b}

\subsubsection{Background}
WASP-39\,b is an inflated warm Neptune orbiting a late G-type star \citep{faedi_wasp-39b_2011}. It is one of the archetype exoplanets for atmospheric characterization and comparison due to its cloud-free high metallicity atmosphere \citep{wakeford_complete_2018,sing_continuum_2016}. Detections of water, carbon monoxide, carbon dioxide, and hydrogen sulfide have been reported with HST, Spitzer, and the newly launched JWST \citep{wakeford_complete_2018,jwst_transiting_exoplanet_community_early_release_science_team_identification_2023,rustamkulov_early_2023, feinstein_early_2023,alderson_early_2023,ahrer_early_2023,tsai_photochemically_2023}. However, no studies have been reported for its upper atmosphere.
\subsubsection{Helium triplet}
The two time series (2022-06-04 and 2022-06-08) respectively cover the full transit with baseline and the transit until the start of egress with baseline only before transit. The weakest components of the OH doublets overlap with the red wing of the stellar helium triplet but are well corrected (Fig.\,\ref{fig:Master_out}). Some features are visible in the transmission spectrum and the helium light curve (Figs.\,\ref{fig:Average_TS} and \ref{fig:Average_LC}) at different wavelengths and phases. We attribute these features to instrumental systematics rather than the presence of helium in the exoplanet atmosphere. To reinforce this point, the bootstrap analysis (Fig.\,\ref{fig:Boot_strap}) shows distributions with no excess absorption for the out-out and in-in scenarios while the in-out scenario mean value varies between the two nights from positive to negative excess absorption value, but is still compatible with no excess absorption. The measured excess absorption on the transmission spectrum is 0.47\,$\pm$\,0.68\,\%, but the 3-$\sigma$ upper limit on the presence of helium is set at $<$2.05\,\%. 

\subsection{WASP-52\,b}

\subsubsection{Background}
WASP-52\,b is a hot Jupiter orbiting around an active K dwarf \citep{hebrard_wasp-52b_2013}. While detection of water and clouds in its lower atmosphere have been reported \citep{alam_hst_2018,tsiaras_population_2018,bruno_wasp-52b_2020}, WASP-52\,b was intensively more studied for its upper atmosphere. Detections of the sodium and potassium doublets and the H-$\alpha$ line in the visible with ESPRESSO \citep{chen_detection_2020} indicate an extended thermosphere above the cloud deck up to $\sim$1.2\,R$_p$, still below the Roche lobe radius (1.75\,R$_p$). These detections are a bit surprising with respect to the equilibrium temperature of $\sim$1200\,K, but could be explained by hot upper layers due to the strong stellar XUV flux correlated to stellar activity. More recently, metastable helium was detected at high-resolution by \cite{kirk_kecknirspec_2022} but only an upper limit was set at low resolution \citep{vissapragada_constraints_2020}. \cite{kirk_kecknirspec_2022}, one of the strongest, excess absorption of 3.44$\pm$0.31\,\% with NIRSPEC, such that helium almost fill the Roche lobe. They, further, applied \texttt{p-winds} \citep{dos_santos_p-winds_2022} to estimate that the planet loses 0.5\% of its mass per Gyr.

\subsubsection{Helium triplet}
The two time-series cover the full transit with a baseline before and after transit. The strong component of the telluric OH line overlaps partially and completely with the helium triplet for the two nights (Fig.\,\ref{fig:Master_out}). We note that the telluric-corrected master-out of each night is not perfectly identical. This is likely due to the low SNR of the datasets. This also impacts the shape of the stellar helium line where it is shallower and broader for the first night. Nonetheless, the transmission spectra and helium light curves of both nights are in good agreement with each other. From the transmission spectroscopic map (Fig.\,\ref{fig:TS_map}), we can see that the before-transit and in-transit spectra have features in excess absorption across the spectral range. This is also captured by the helium light curve (Fig.\,\ref{fig:Average_LC}). In the averaged transmission spectrum (Fig.\,\ref{fig:Average_TS}), noise structures are also visible, but no significant excess absorption is detected at the helium line position and it is concurred by the bootstrap analysis (Fig.\,\ref{fig:Boot_strap}) of the two time-series. The measured excess absorption on the transmission spectrum is 1.36\,$\pm$\,0.56\,\%, and the 3-$\sigma$ upper limit on the presence of helium is set at $<$1.69\,\%. From the Allan plot (Fig.\,\ref{fig:Allan}), the WASP-52b data best follow the white noise estimations even if features have large amplitudes up to 4\,\% they are below the 3-$\sigma$ upper limit once integrated over the 0.75\,\AA\ passband. This is in strong disagreement with the results reported by \cite{kirk_kecknirspec_2022} and \cite{vissapragada_upper_2022}. We note that the value reported by \cite{kirk_kecknirspec_2022} is the maximum absorption of their signature, but once integrated over a 0.75\,\% passband their absorption is $\sim$2.7\,\%, which is still in strong disagreement with our observations. We also performed an injection-recovery test by adding to our transmission spectrum a Gaussian of amplitude 3.44\,\% and FWHM of 1\,\AA\ following the result of \cite{kirk_kecknirspec_2022}. The measured excess absorption on this injected data is 3.93\,$\pm$\,0.56\% (7$\sigma$), which confirm the tension with  the literature. Here again, we need more data to settle the debate on the presence of metastable helium.

\subsection{WASP-69\,b}

\subsubsection{Background}
WASP-69\,b is a warm Neptune orbiting an active K dwarf \citep{anderson_three_2014}. It is one of the best targets for atmospheric characterization due to its large-scale height and was, therefore, well studied at high spectral resolution and combined with data at low resolution. \cite{guilluy_gaps_2022} reported the presence in the lower atmosphere of five molecules at more than 3$\sigma$ with the near-infrared high-resolution spectrograph GIANO, but with variability for some molecules between transits. Nonetheless, water was independently confirmed with HST \citep{tsiaras_population_2018,estrela_detection_2021} alongside aerosols. The detection of the sodium doublet in the thermosphere was also reported at high spectral resolution \citep{casasayas-barris_detection_2017,khalafinejad_probing_2021}, but with a strong amplitude ratio between the two lines, which is likely due to the presence of hazes. WASP-69\,b is also one of the two first exoplanets (with HAT-P-11\,b, \citealt{allart_spectrally_2018}) with a measured excess absorption of helium obtained at high resolution with CARMENES \citep{nortmann_ground-based_2018}. The authors measured a blueshifted line profile with a clear excess absorption of 3.59$\pm$0.19\% with a slight excess after the opaque transit. This is in agreement with an extended thermosphere up to 2.2\,R$_p$. This signature was also confirmed at low resolution by \cite{vissapragada_constraints_2020}.
\subsubsection{Helium triplet}
The time series of WASP-69\,b covers the full transit with a baseline before and after. The strong component of the OH lines overlaps with the redwing of the helium triplet (Fig.\,\ref{fig:Master_out}). A clear signature is visible during the transit slightly blue-shifted to the expected position of the helium triplet, but can still be associated with both the stellar and planetary rest frame (Fig.\,\ref{fig:TS_map}, \ref{fig:Average_TS} and \ref{fig:Average_LC}). From the helium light curve, it is not possible to confirm the presence of post-transit absorption as discussed in \cite{nortmann_ground-based_2018}. Moreover, we see that there is some variability along the transit duration with a maximum of excess absorption before mid-transit. The measured excess absorption on the transmission spectrum is 2.21\,$\pm$\,0.46\,\% (4.8\,$\sigma$) with a maximum excess absorption of $\sim$3.1\,\%.  This is significantly below the reported maximum excess absorption measured by \cite{nortmann_ground-based_2018} but the integrated signal over a 0.75\,\AA\ bandpass is in agreement with an absorption of $\sim$2\,\% as their line profile is quite narrow. The difference at the maximum of excess absorption is too large to be explained by data reduction effects but could be caused by some instrumental or systematic effects as well as some astrophysical variability linked either to the star or the planet.

\subsection{WASP-80\,b}

\subsubsection{Background}
WASP-80\,b is a hot Jupiter orbiting a K7 dwarf \citep{triaud_wasp-80b_2013}. Broadband absorption features of water and carbon dioxide partly muted by clouds and aerosols reveal an enhanced atmospheric metallicity \citep{wong_hubble_2022}. No detection of metastable helium has been reported either at low resolution by \cite{vissapragada_upper_2022} nor at high resolution by \cite{fossati_gaps_2022}. The latter set an upper limit of 0.7\% was set with GIANO with 3 transits. The authors estimated that the helium-to-hydrogen abundance ratio of WASP-80\,b has to be lower than solar to match their data.
\subsubsection{Helium triplet}
The time series of WASP-80\,b observed with SPIRou spans the full transit with a baseline before and after. The telluric line positions overlap with the helium triplet with the strong component of the OH doublet on the bluewing and the water telluric line on the redwing (Fig.\,\ref{fig:Master_out}). The telluric (absorption and emission) lines seem to be well corrected and should not impact the potential presence of planetary helium. Systematics are present in the transmission spectrum and the helium light curve (Figs.\,\ref{fig:Average_TS} and \ref{fig:Average_LC}), but are not related to the presence of helium in the exoplanet atmosphere. The bootstrap analysis (Fig.\,\ref{fig:Boot_strap}) shows distributions with no excess absorption for the out-out, in-in, and in-out scenarios. The measured excess absorption on the transmission spectrum is 0.03\,$\pm$\,0.41\,\%. The 3-$\sigma$ upper limit on the presence of helium was set at $<$1.24\,\%, which is less stringent than the upper limit set by \cite{fossati_gaps_2022} even if scaled to 3 transits and with the same upper limit metric.

\subsection{WASP-127\,b}

\subsubsection{Background}
WASP-127\,b is a bloated hot Neptune on a misaligned circular orbit around an old ($\sim$10\,Gyr) G-type star \citep{lam_dense_2017,allart_wasp-127b_2020}. With its large scale height, WASP-127\,b is one of the most amenable planets for atmospheric characterization. \cite{spake_abundance_2021} indeed revealed with HST and Spitzer a feature-rich atmosphere with the strongest amplitude known ($\sim$800\,ppm) for the water band at 1.4\,$\mu$m. In addition, they constrained the presence of clouds, aerosols, and of carbon-bearing species without the possibility to distinguish between a CO-rich high C/O ratio atmosphere or a CO$_2$-rich low C/O ratio atmosphere. High-resolution observations with SPIRou \cite{boucher_co_2023} reported the detection of water and a possible hint of OH, but did not detect the presence of CO. By combining their data with the data of \cite{spake_abundance_2021}, their model tends to favor the low C/O atmosphere. Although the presence of many species in the thermosphere could have been expected, only sodium was detected with ESPRESSO \citep{allart_wasp-127b_2020}, which extends over 7 scale heights only and strong upper limits were set for the potassium doublet and H-$\alpha$. Similarly, \cite{dos_santos_search_2020} reported an upper limit of 0.87\% on the presence of metastable helium with one transit with Gemini/Phoenix, which is probably due to the relatively mild high-energy environment around the star.
\subsubsection{Helium triplet}
Due to the long transit duration of WASP-127b, the three time-series do not cover the full transit and have a little baseline before or after. Only for the last time series (2021-05-03), there is an overlap between the strong component of the OH doublet with the redwing of the stellar helium position (Fig.\,\ref{fig:Master_out}). We note that the stellar helium line is broad and shallow. The transmission spectroscopy map (Fig.\,\ref{fig:TS_map} and \ref{fig:TS_map_night}) exhibits noise structures in the observer or stellar rest frame during the transit, which impacts the transmission spectrum (Fig.\,\ref{fig:Average_TS}). They might be caused by instrumental systematics not caught by \texttt{APERO} or caused by small telluric residuals from the weak component of the OH lines. The average transmission spectrum has a broadband slope that we detrend with a polynomial of order 2. The helium light curve (Fig.\ref{fig:Average_LC}) does not show any excess absorption during transit, which is confirmed by the bootstrap analysis of the in-out scenario (Fig.\,\ref{fig:Boot_strap}). We note the unusual trimodal distribution of the out-out scenario for the first and last night (2020-03-11 and 2021-05-03), which is likely caused by the lack of out-of-transit spectra. The measured excess absorption on the transmission spectrum is 0.05\,$\pm$\,0.16 \,\%. We put a 3-$\sigma$ upper limit on the presence of helium at $<$0.48\,\%, which improves the previous constraint set by \cite{dos_santos_search_2020}.

\begin{table*}
	\centering
	\caption{Summary of the helium triplet measurements from this work.}
	\label{tab:results}
	\begin{tabular}{cccccccc} 

		\hline
		Target 			& Excess Absorption		& Eq. opaque radius 		& H 			& $\delta$R$_p$/H	& $\mathrm{F_{5-504\mathring{A}}} $ 					& Roche Lobe  	&$\dot{m}$ \\

  		Units			& \%						&  R$_p$					&  km		&					& $\mathrm{10^4\cdot erg\cdot s^{-1}\cdot cm^{-2}}$ 	& R$_p$			&$\mathrm{10^{11}\cdot g\cdot s^{-1}}$  \\

		\hline
		AU\,Mic\,b 	 	& $<$0.26 	 			& $<$ 1.67 				&  399 	 	& $<$ 38 	 		&  158.50 											& 7.7 			&	$<$ 1.51	\\
		GJ1214\,b 	 	& $<$2.92 	 			& $<$ 1.77 				&  339 	 	& $<$ 40 	 		&  1.97 												& 3.2 			& 	--				\\	
		GJ3470\,b 	 	& $<$ 0.64 	 			& $<$ 1.36 	 			&  666 	 	& $<$ 16 	 		&  2.97 												& 3.1 			&	$<$ 1.41	\\
		HAT-P-11\,b 		& 0.76$\pm$0.07 			& 1.80$\pm$0.05 	 		&  352 	 	& 63$\pm$4  			&  0.16												& 6.5 			&	0.67$^{+0.27}_{-0.24}$		\\
		HD189733\,b 		& 0.69$\pm$0.04 			& 1.14$\pm$0.01 	 		&  331 		& 33$\pm$2 			&  0.42												& 3.0 			&	0.94$^{+0.82}_{-0.61}$		\\
		WASP-11\,b 	 	& $<$ 1.56	 			& $<$ 1.38 	 			&  484 	 	& $<$ 55   			&  0.19												& 3.7 			&	$<$ 0.08	\\
		WASP-39\,b 	 	& $<$2.04 				& $<$ 1.43 	 			&  1635	 	& $<$ 24	 			&  0.12												& 2.6 			&	$<$ 1.50	\\
		WASP-52\,b 	 	& $<$ 1.69				& $<$ 1.27 				&  1114		& $<$ 22 	 		&  20.93 											& 1.7 			&	$<$ 6.99  \\
		WASP-69\,b 	 	& 2.35$\pm$0.46 			& 1.50$\pm$0.09 			&  1001		& 38$\pm$7 			&  0.93												& 2.9 			& 	0.40$^{+0.58}_{-0.25}$\\
		WASP-80\,b 	 	& $<$ 1.24 				& $<$ 1.19 	 			&  373 	 	& $<$ 35 	 		&  31.78												& 3.4 			&	$<$ 0.14	\\
		WASP-127\,b 		& $<$ 0.48 	 			& $<$ 1.20 				&  3762  	& $<$ 5 	 			&  0.12												& 2.0 			&	--				\\
		\hline
	\end{tabular}
\tablefoot{Excess absorption, equivalent radius, and $\delta$R$_p$/H are the 3-$\sigma$ upper limits unless a detection is reported. F$_{5-504\mathring{A}}$ is the integrated XUV flux from 5 to 504\,\AA\ and scaled to the semi-major axis of the planets. The Roche Lobe was estimated following \citet{rappaport_roche_2013}.}
\end{table*}

\section{Interpretation} \label{sec:interpretation}
\subsection{Stellar pseudo-signal}
During a transit, the planet occults different stellar regions where metastable helium can be present or not. This can imprint the transmission spectrum with an absorption or emission spectral feature mimicking planetary signals. We studied the impact of a stellar pseudo-signal with the model described in section\,\ref{model:pseudosignal} for the following planets:  AU\,Mic\,b, GJ\,1214\,b, GJ\,3470\,b, HAT-P-11\,b, HD\,189733\,b, WASP-52\,b, and WASP-69\,b, which are the planets where a stellar pseudo-signal could play a role on the presence of helium or its variability. For all the systems, we explored the impact of the filling factor, $f$, with values between 0.2 and 1 on the strength of the stellar pseudo-absorption signal but no significant variations were measured. Table\,\ref{tab:pseudosignal} reports the maximum integrated pseudo-signal excess absorption for the different planets assuming the planets only transit bright regions and all the stellar helium absorption comes from dark regions ($\alpha$=0). For  AU\,Mic\,b, GJ\,1214\,b, GJ\,3470\,b, and HAT-P-11\,b, the impact of stellar pseudo-signal is negligible due to their small $R_P/R_{\star}$ and cannot explain the measured excess absorptions or their variability. However, we note that in the case of GJ\,3470\,b, a pseudo-emission signal could reduce the helium signature amplitude by $\sim$0.6\,\% if the planet was only transiting dark regions ($f$=0.75 based on \cite{andretta_estimates_2017} and a helium EW of $\sim$270\,m\AA) during our observations, and could partly explain the observed discrepancy with \cite{palle_he_2020}.\\
In the case of the larger planets, namely HD\,189733\,b, WASP-52\,b, and WASP-69\,b, the maximum pseudo-absorption signal from the star can contribute to the variability observed between different transits but cannot be the single cause of the He absorption seen and other processes are required. 

\begin{table}
	\centering
	\caption{Strength of the maximum pseudo-absorption signal producible}
	\label{tab:pseudosignal}
	\begin{tabular}{cc} 

		\hline
		Target 				& Pseudo-signal excess absorption [\%]  \\
		\hline
		AU\,Mic\,b 	 	& $\sim$0.03 	 				\\
		GJ\,1214\,b 	 	& $\sim$0.05					\\	
		GJ\,3470\,b 	 	& $\sim$0.21					\\
		HAT-P-11\,b 		& $\sim$0.02	 				\\
		HD\,189733\,b 	& $\sim$0.75	 				\\
		WASP-52\,b 	 	& $\sim$0.95				 	\\
		WASP-69\,b 	 	& $\sim$0.53	 				\\
		\hline
	\end{tabular}
\end{table}

We describe in the following two subsections the extreme case of HD\,189733\,b where the pseudo-signal can have a similar excess absorption than the observed one and the case of HAT-P-11b as the best target to study planetary variability.
\subsubsection{HD\,189733\,b: impact of stellar variability}
\begin{figure}
\includegraphics[width=\columnwidth]{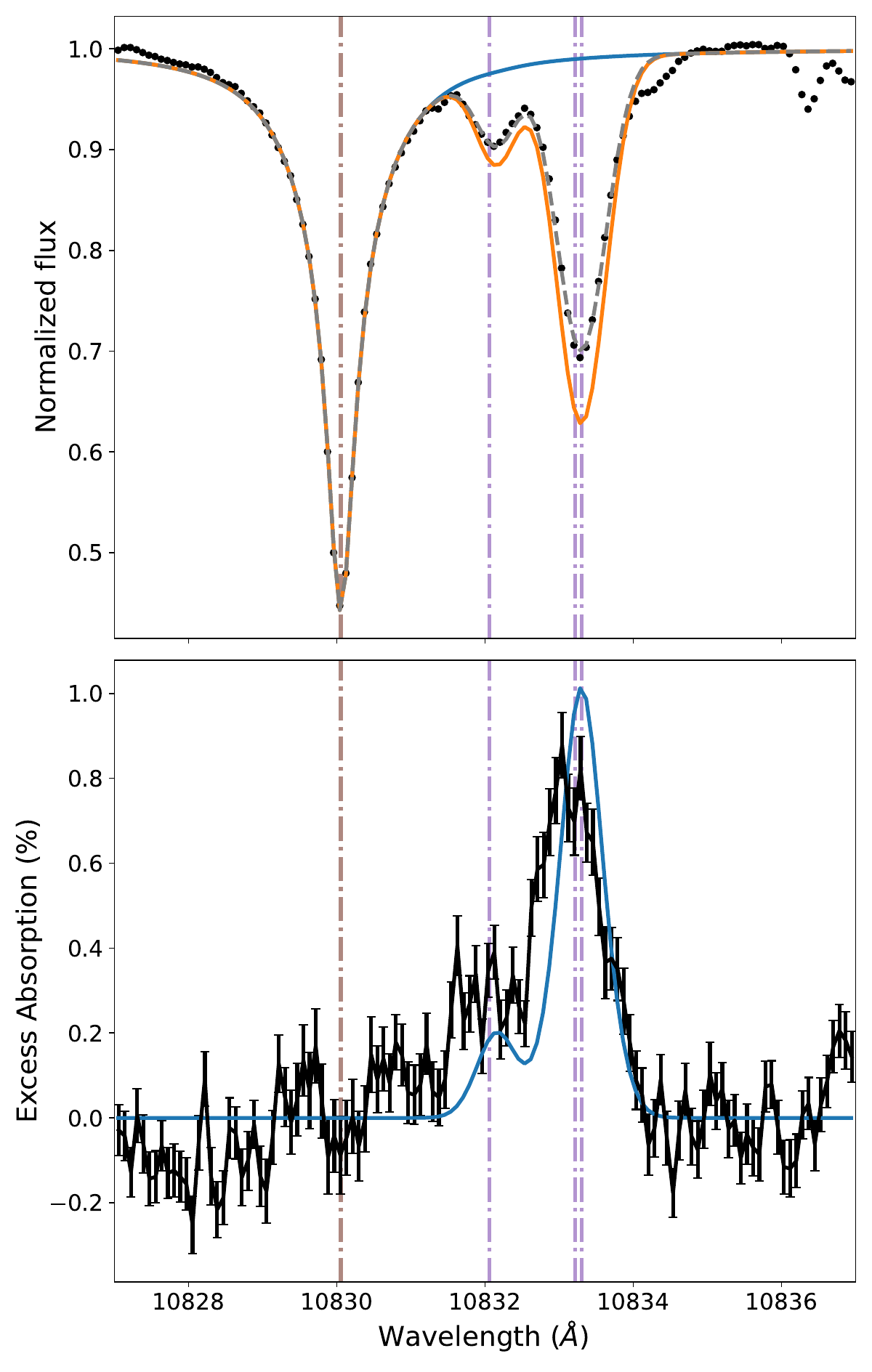}
\caption{Impact of stellar pseudo-signal on the transmission spectrum of HD\,189733\,b.  \textit{Top:} Stellar master spectrum (black dots) with the modeled stellar spectra of bright (orange) and dark (blue) regions and both combined (dashed grey) for a filling factor of 0.8. \textit{Bottom:} Observed transmission spectrum in black. Average transmission spectrum caused by stellar pseudo-absorption (blue) signal. The dashed-dotted brown line is at the position of the \ion{Si}{i} line and the dashed-dotted purple lines are at the position of the \ion{He}{i} lines in the star (\textit{top}) and planet (\textit{bottom}) rest frame.}
    \label{fig:PS_HD189733}
\end{figure}

The measured helium EW is $\sim$295\,m\AA\, which is very close to the value measured by \cite{salz_detection_2018}, and sets the filling factor value to 80\% \citep{andretta_estimates_2017}. As it was discussed in \cite{salz_detection_2018} and \cite{guilluy_gaps_2020}, the impact of a stellar pseudo-signal is significant for HD\,189733\,b (Fig.\,\ref{fig:PS_HD189733}) due to its large $R_P/R_{\star}$ with a value of $\sim$0.75\%, which is equivalent to the measured excess absorption on the transmission spectrum. However, it is important to note that the line shape of the pseudo-signal does not match our measured helium line shape. On the latter, there is a clear additional blueshifted signal that cannot be reproduced by pseudo-signal. In addition, the strength of the pseudo-signal slightly overestimates the observed one and requires the production of helium in the bright region as well to decrease it. Therefore, it is not possible that all the detected signal is of stellar origin, such that a significant fraction must come from the planet's atmosphere. However, the observed variability between transits (and instruments) might come from stellar variability. 

\subsubsection{HAT-P-11b: the advantage of its eccentric orbit}
\begin{figure}
\includegraphics[width=\columnwidth]{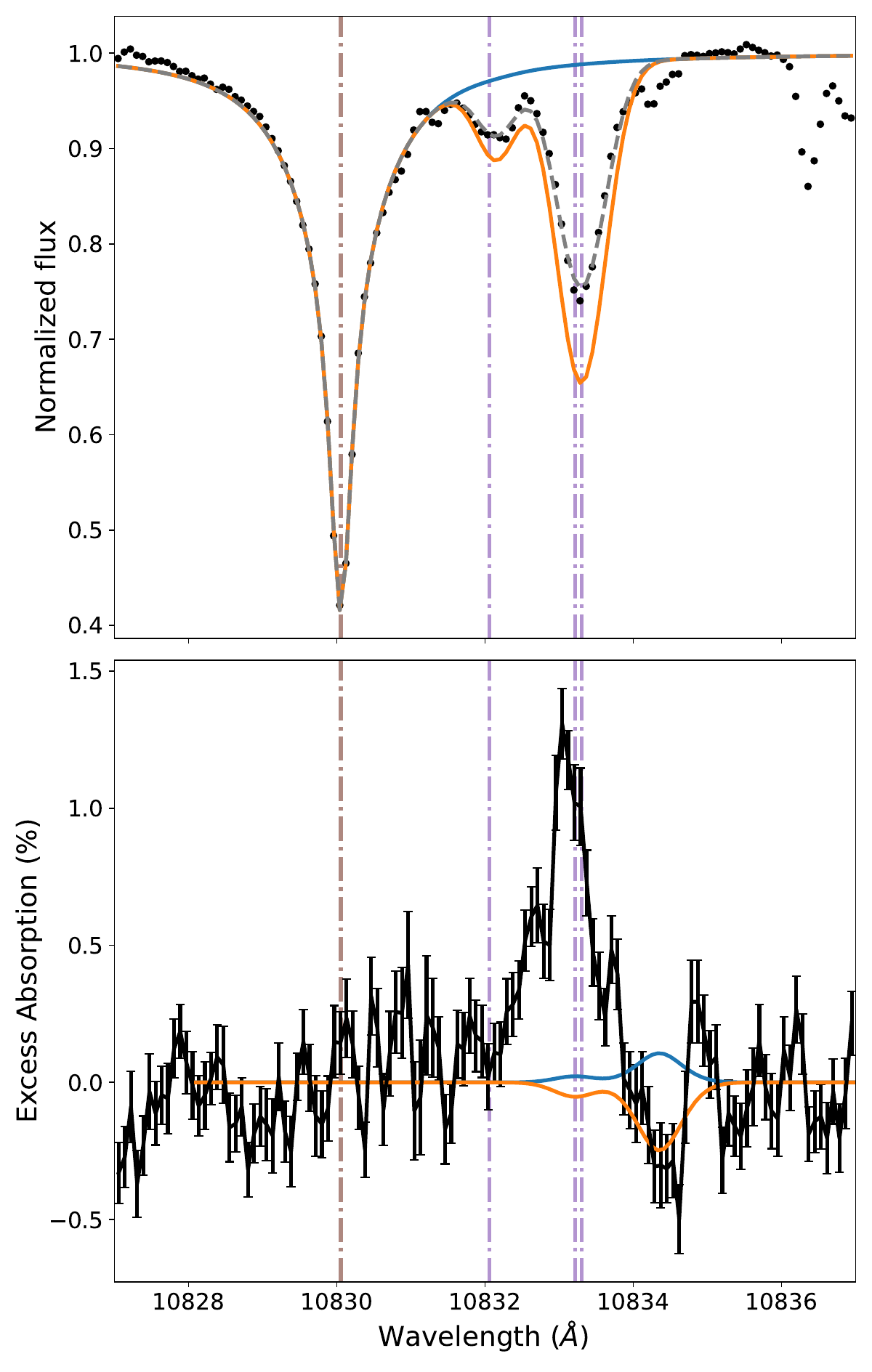}
\caption{Same as Fig.\,\ref{fig:PS_HD189733} for a filling factor of 0.7. \textit{Bottom:} Average transmission spectra caused by stellar pseudo-absorption (blue) and pseudo-emission (orange) signals.}
    \label{fig:PS_HATP11b}
\end{figure}

Based on \cite{andretta_estimates_2017} and an EW of the stellar helium line of $\sim$240\,m\AA, we estimate a filling factor of $\sim$0.7. As shown in Fig.\,\ref{fig:PS_HATP11b}, the modeled stellar pseudo signal contributes $\sim$0.02\,\% to the absorption measured at the positions of the helium lines  in the planet rest frame. This is less than the 1$\sigma$ uncertainty and is due to the eccentric orbit of HAT-P-11\,b, which decorrelates the planetary track from the stellar one. This strengthens the planetary origin scenario for the slight variability of the helium triplet between transits. Interestingly, the impact of stellar pseudo-signal could explain the feature visible on the red wing of the helium triplet in \cite{allart_spectrally_2018} in absorption and here in emission depending on the occultation of bright or dark regions. 

\subsection{Atmospheric modeling}
\begin{figure*}
\centering
\includegraphics[width=0.92\textwidth]{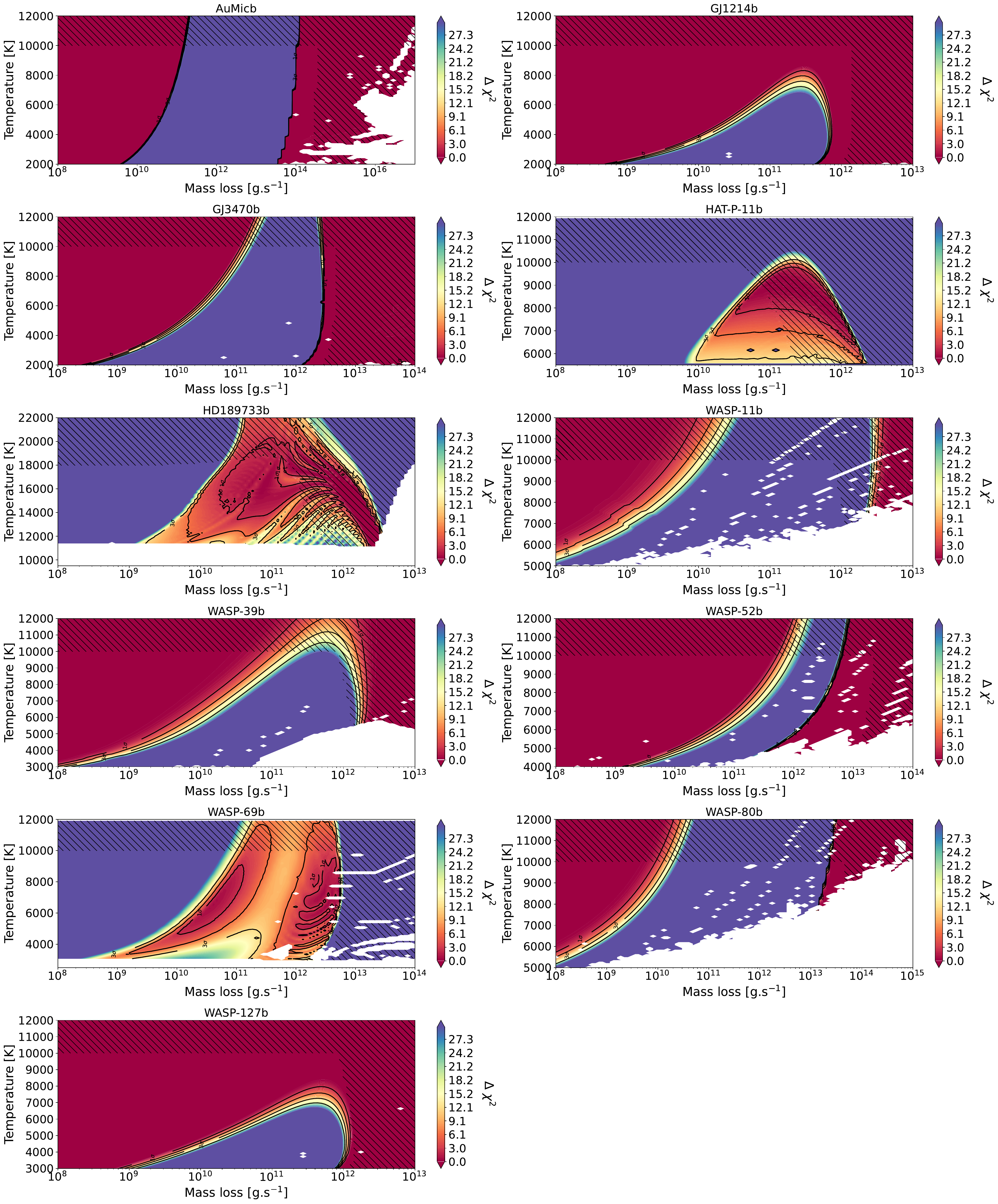}
\caption{$\Delta\chi^2$ maps of mass-loss rate and temperature for the eleven planets. Regions of the parameter space in red are in agreement with the data, while models in blue cannot reproduce the observed transmission spectra. The hatched region of the parameter space is physically excluded.}
\label{fig:chi2map}
\end{figure*}

Figure\,\ref{fig:chi2map} shows the $\Delta\chi^2$ of the atmospheric best fit model as a function of mass-loss rate and temperature of the thermosphere as derived from the \texttt{p-winds} models \citep{dos_santos_p-winds_2022}. Regions of the parameter space in red are in agreement with the data, while models in blue cannot reproduce the observed transmission spectra. Some simulations did not converge properly at each altitudes due to numerical issues in \texttt{p-winds} related to high variation of the different contributions. It is consequently better to exclude them and they are represented as white regions. The hatched region of the parameter space is physically excluded based on the gravitational potential for temperature and on the energy limited for the mass-loss rate (see section\,\ref{model:pwinds}). \\
For the non-detections, we are able to identify regions of the parameter space that are in disagreement with the data but with a clear correlation between temperature and mass-loss rate. However, these regions are split between models with low mass-loss rate and high mass-loss rate that both are within the data error bars. This contradiction can be explained by our choice of homogeneous analysis and fixed thermosphere radius to the Roche lobe. The metastable helium profiles for the models at high mass-loss rates have a large fraction of helium particles at altitudes higher than the Roche Lobe. Consequently, the quantity of metastable helium below this radius is sufficient to reproduce the non-detections observed. As described before, we do not consider these models as realistic and therefore used the 3-$\sigma$ contour of the regions of the parameter space at a low mass-loss rate to set its upper limit (Table\,\ref{tab:results}). However, it was not possible to derive upper limits for GJ\,1214\,b, and WASP-127\,b as there is always one model at a given temperature that fit the data independently of the mass-loss rate. We note that the upper limit on the mass-loss rate of GJ\,3470\,b is similar to the one derived by \cite{palle_he_2020} of 3$\cdot$10$^{10}\,\mathrm{\,g\cdot s^{-1}}$. However, we constrained less well the mass-loss rate of WASP-52\,b compared to \cite{kirk_kecknirspec_2022}.\\
For the detections, the thermospheric radius, where the \texttt{p-winds} simulation is stopped, was set as a free parameter with an upper limit at the Roche Lobe along a free line-of-sight bulk velocity. We allow the thermospheric radius to be below the Roche Lobe radius in case of a compact thermosphere probed by the helium triplet. The \texttt{p-winds} best-fit models are displayed in Fig.\,\ref{fig:Average_TS} for the three detections: HAT-P-11\,b, HD\,189733\,b, and WASP-69\,b, while the $\Delta\chi^2$ map of mass-loss rate and temperature is shown in Fig.\,\ref{fig:chi2map}. \\
The best-fit model for HAT-P-11\,b is obtained for $\dot{m}$=0.67$^{+0.27}_{-0.24}\cdot10^{11}\,\mathrm{\,g\cdot s^{-1}}$, T=8726$^{+158}_{-557}$\,K, $v$=$-$5.3$\pm$0.8$\,\mathrm{km\cdot s^{-1}}$ and $r$= 6.5$^{+0}_{-1.5}\,\mathrm{R_p}$. We confirm the blueshifted nature of the helium triplet reported by \cite{allart_spectrally_2018} ($v\sim$-3$\,\mathrm{km\cdot s^{-1}}$), which is marginally at 2-3 $\sigma$. Our results can be compared to \cite{dos_santos_p-winds_2022}, where the authors benchmarked the use of \texttt{p-winds} on the HAT-P-11\,b data of \cite{allart_spectrally_2018} obtained with CARMENES. The only caveat is that we let the radius free, but it cannot be higher that the limit set at the Roche Lobe, indicative of an exospheric contribution. Nonetheless, we find a good agreement with \cite{dos_santos_p-winds_2022} for both the temperature and the mass-loss rate. The comparison with the results of \cite{allart_spectrally_2018} is less straightforward as the authors used the 3D code EVE that simulates both the thermosphere and exosphere, which is more complex than a Parker wind model. For example, we derive a lower temperature but a higher mass-loss rate. This shows that the derivation of the physical parameters of the thermosphere highly depends on the models used and their assumptions. \\
The best-fit model for HD\,189733\,b is obtained for $\dot{m}$=0.94$^{+0.82}_{-0.60}\cdot10^{11}\,\mathrm{\,g\cdot s^{-1}}$, T=16690$^{+1966}_{-2182}$\,K, $v$=$-$4.2$\pm$0.8$\,\mathrm{km\cdot s^{-1}}$ and $r$= 1.41$^{+0.20}_{-0.03}\,\mathrm{R_p}$. Due to limitations of the \texttt{p-winds} code\footnote{Some differential equations cannot be solved depending of the initial parameters.}, it was not possible to compute models for temperatures lower than 11\,320\,K, which reduced the explored parameter space. The measured blueshift is in agreement at 1-$\sigma$ with the values of \cite{salz_detection_2018} and \cite{guilluy_gaps_2020}. We confirm that the region of HD\,189733\,b atmosphere with helium in the triplet state is hot, compact, and sizes to only $\sim$0.2\,$\mathrm{R_p}$ \citep{salz_detection_2018,guilluy_gaps_2020,lampon_modelling_2021}. The derived mass-loss rate is similar to \cite{lampon_modelling_2021} but they assumed an almost fully ionized atmosphere with a very low mean molecular weight of H/He=99.2/0.8, which was necessary to fit their data.\\
The best-fit model for WASP-69\,b is obtained for $\dot{m}$=0.40$^{+0.58}_{-0.25}\cdot10^{11}\,\mathrm{\,g\cdot s^{-1}}$, T=6987$^{+1617}_{-1604}$\,K, $v$=$-$5.4$\pm$1.2$\,\mathrm{km\cdot s^{-1}}$ and $r$= 2.9$^{+0}_{-0.9}\,\mathrm{R_p}$. We confirm the blueshifted nature of the helium triplet reported by \cite{nortmann_ground-based_2018} ($v$=$-$3.58$\pm$0.23$\,\mathrm{km\cdot s^{-1}}$) compatible at 2-$\sigma$ and higher velocity. The derived mass-loss rate is also consistent with 3D hydrodynamics and self-consistent photochemistry models of \cite{wang_metastable_2021}. The best-fit model uncertainty on the radius indicates that a radius larger than the Roche Lobe could be preferred but it requires the use of more complex models to describe the exosphere \citep[e.g.,][]{bourrier_atmospheric_2013, allart_spectrally_2018, allart_high-resolution_2019,allan_evolution_2019, wang_metastable_2021}.

\section{Discussion} \label{sec:discussion}
\begin{figure*}
\includegraphics[width=\textwidth]{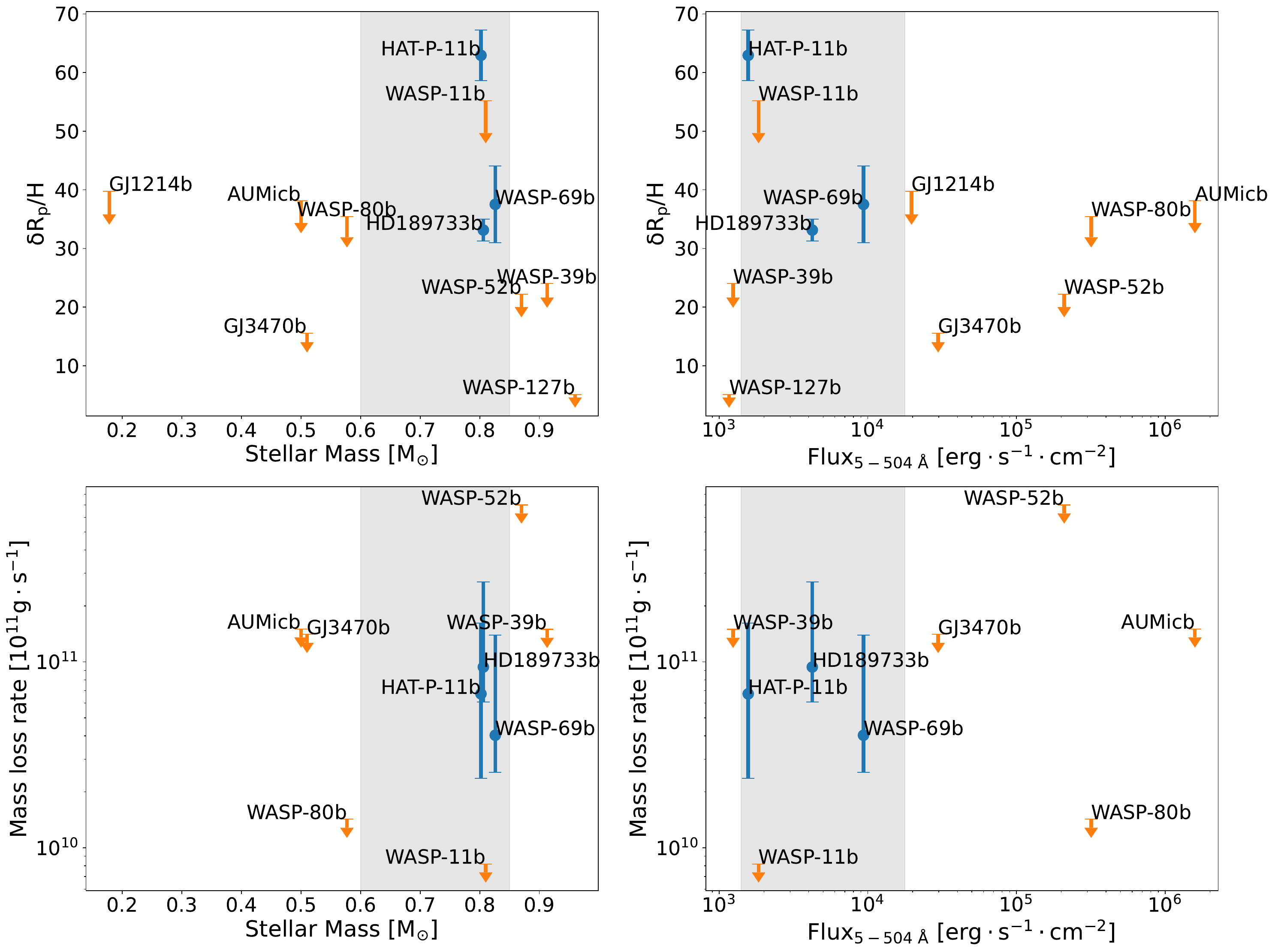}
\caption{Correlation plots for the eleven planets studied in this paper. Detections are shown as blue points and upper limits as orange points. We pinpointed with a grey area the parameter space that seems to favor the detection of metastable helium in exoplanet atmospheres. \textit{Top left:} $\delta$R$_p$/H as a function of the stellar mass. \textit{Top right:} $\delta$R$_p$/H as a function of XUV flux received from 55 to 5504\,\AA. \textit{Bottom left:} mass-loss rate derived from the \texttt{p-winds} simulations as a function of the stellar mass. \textit{Bottom right:} mass-loss rate derived from the \texttt{p-winds} simulations as a function of XUV flux received from 5 to 504\,\AA.}
    \label{fig:Correlations}
\end{figure*}

From the upper limits and detections measured in section\,\ref{sec:survey}, we estimated the equivalent opaque radius, which can be normalized by the planet scale height to produce the quantity $\delta$R$_p$/H proposed in \cite{nortmann_ground-based_2018}. This quantity expresses the number of scale heights that is probed by the helium triplet in the exoplanet atmosphere. We assumed here the equilibrium temperature and a mean molecular weight of 1.3 to estimate the scale height. We explored how $\delta$R$_p$/H correlates with various system parameters: stellar mass, stellar radius, effective temperature, age, planetary mass, planetary radius, planetary density, and equilibrium temperature. We also extended the search for possible correlations with the stellar XUV flux scaled to the semi-major axis of the planets measured between 5 and 504\,\AA\, which is the part of the XUV flux responsible for the creation of metastable helium in exoplanet atmospheres \citep{sanz-forcada_active_2008}. However, these flux values are model-dependent and are subject to various assumptions such as the age of the system (see section\,\ref{model:pwinds}). We limited the search for correlation to the sample studied here because the data were obtained with the same instrument and reduced in a homogeneous way including the report of detections and upper limits, which differs in the literature from one paper to another. We note that a trend is noticeable with stellar age (not shown), but due to the lack of precision on the stellar age and the small number of targets in our sample, it is not possible to draw more conclusions. \\
The top panels of Fig.\,\ref{fig:Correlations} show the correlations for $\delta$R$_p$/H with the stellar mass and the XUV flux, the two parameters showing the strongest trends. We note that the upper limit set for WASP-11\,b is not constraining enough to be useful. However, more observations of this planet might reveal the presence of metastable helium as the system is quite similar to the reported detections. The correlation with the stellar mass is well identified where the presence of metastable helium around exoplanet is favored for planets orbiting stars with masses between $\sim$0.6 and $\sim$0.85\,M$_{\odot}$, which corresponds to K dwarfs as predicted by \cite{oklopcic_helium_2019}. This range of stellar mass also agrees with previous detections and non-detections published in \cite{kasper_nondetection_2020, casasayas-barris_carmenes_2021, zhang_escaping_2022,zhang_detection_2022} for example. However, it is in contradiction with the detection of helium obtained for HD\,209458\,b by \cite{alonso-floriano_he_2019} or for HAT-P-32b by \cite{czesla_h_2022} for which the stellar masses are $\sim$1.2\,M$_{\odot}$. We can also identify a range of XUV flux received by the planets that seems to favor the presence of metastable helium between 1\,400 and 17\,800\,erg$\,\cdot\,$s$^{-1}\,\cdot$\,cm$^{-2}$. Nethertheless, as discussed above, these XUV values are model-dependent and linked to the stellar ages, which is usually not well constrained. We see that on the one hand WASP-39\,b and WASP-127\,b receive less XUV flux while orbiting G-type stars, and are the oldest planets studied here. On the other hand  AU\,Mic\,b, WASP-52\,b, and WASP-80\,b are the youngest planets and receive the highest amount of XUV flux. It is also interesting to note that WASP-52\,b, for which we have a contradictory result with \cite{kirk_kecknirspec_2022}, is well above the favored range of XUV flux range even with a well-constrained age, which is in contrast with its proximity to the acceptable range of stellar mass that seems to favor the presence of metastable helium. \\
The bottom panels of Fig.\,\ref{fig:Correlations} show the correlations for $\dot{m}$ with the stellar mass and the XUV flux. The use of $\dot{m}$ should be preferred to the $\delta$R$_p$/H as it is a more physical quantity to describe the thermosphere probed by the helium triplet. Indeed, the $\delta$R$_p$/H quantity assumes H and He particles in their neutral state only and at the equilibrium temperature of the planet, which is much lower than the thermospheric temperature. However, the correlations between $\dot{m}$ with stellar mass and XUV flux are not well defined as upper limits on $\dot{m}$ are in agreement with some of the derived $\dot{m}$ for detections. This can be linked to the correlation reported in section\,\ref{sec:interpretation} between $\dot{m}$ and $T$. We note that the best upper limits on $\dot{m}$ are for WASP-11b and WASP-80b, while they have poorly constrained $\delta$R$_p$/H as opposed to AU\,Mic\,b, GJ\,3470\,b, WASP-39\,b and WASP-52\,b. This is surprising for WASP-11\,b as all the planets within the same stellar mass and XUV flux range have clear detections and higher $\dot{m}$.
Interestingly, based on their gravitational potential all the planets studied here are expected to fall in the strong hydrodynamic wind regime (intermediate regime for HD\,189733\,b ) and thus undergo strong evaporation \citep{salz_simulating_2016}. However, we do not observe signatures of the helium triplet for most of our targets. This discrepancy already reported by \cite{fossati_gaps_2022} or \cite{vissapragada_upper_2022} could be the result of more complex mechanisms not integrated in 1D hydrodynamical codes. Although, a simpler explanation can be found in the population of the metastable helium triplet. The strength of those helium lines does not depend on the evaporation rate but on the mid-UV flux ionizing the metastable helium particles and on the EUV flux populating the triplet state through recombination. Based on the metastable helium population mechanisms, \cite{oklopcic_helium_2019} suggested that planets orbiting K type stars and receiving the right balance of mid-UV and EUV flux are the most amenable to probe the evaporation through the helium triplet. To summarize, even if for the planets with non detection of the helium triplet strong evaporation can happen, but this process is not traced by the helium triplet as most of the helium particles are in the ground state.

\section{Summary and conclusion}\label{sec:conclusion}
This paper presents the first homogeneous analysis of the metastable helium triplet for eleven exoplanets observed with a single high-resolution near-infrared spectrograph, SPIRou. We confirmed detection of He triplets in the atmosphere of HAT-P-11\,b, HD\,189733\,b and WASP-69\,b. We obtained upper limits for GJ\,3470\,b and WASP-52\,b, that disagree with previously published papers. We set new or confirm upper limits for  AU\,Mic\,b, WASP-11\,b, WASP-39\,b, WASP-80\,b, and WASP-127\,b. We finally obtained an upper limit for GJ\,1214\,b, which is not constraining enough to settle the debate on the presence of helium. \\
We note that the SPIRou transmission spectra are affected by various systematics that are difficult to understand and to properly remove. We mitigated them by scaling our uncertainties but more robust approaches could be consider by combining Gaussian processes with model fitting algorithm (out of the scope of this paper). They can be caused by the low data quality (e.g., close to readout regime), instrumental effects, reduction pipeline errors, or stellar variability like in the case of  AU\,Mic\,b. Nonetheless, we set 3-$\sigma$ upper limits which are as representatives as possible of these systematics and assuming a given helium line width.\\
We estimated the impact of stellar-pseudo signal on the observed helium features with a simple toy model. We concluded that none of the detections could solely be explained by such an effect, but that could contribute to some of the variability observed between transits and instruments. A more complex model would be needed to take into account the complexity of stellar surfaces and their occultation by planets combined with the intrinsic variability of the stellar flux in various wavelength domains. To better understand the impact of stellar variability on the presence of metastable helium in exoplanet atmospheres, applying homogeneous analysis of multiple stellar and planetary tracers (e.g., Na, H-$\alpha$, He) as presented in \cite{guilluy_gaps_2020,czesla_h_2022} will be necessary for the future. Instruments like CARMENES, GIANO simultaneously to HARPS-N, SPIRou simultaneously to ESPaDOnS (in a near future)  and NIRPS simultaneously to HARPS will have an edge to disentangle multi-source effects. Among the three detections, HD\,189733\,b is probably the one most impacted by stellar variability and requires specific analysis to properly extract the true planetary signature. However, HAT-P-11\,b emerges as the best candidate to search for temporal planetary variability signature as it is completely free of stellar contamination, and variability in the CARMENES data \citep{allart_spectrally_2018} still have to be explained.\\
The transmission spectra of the 11 planets were modeled with \texttt{p-winds} \citep{dos_santos_p-winds_2022}. We excluded models at high temperatures and mass-loss rates to stay in a physical thermosphere assumption based on the gravitational potential of the planet and the maximum mass-loss efficiency for a photoionization-driven isothermal Parker wind \citep{vissapragada_maximum_2022}. We also fixed the radius of the thermosphere to the Roche Lobe for the non-detections to derive reasonable constraints. However, we note that discussions on the criteria to set the radius are missing in the literature. Upper limits on the mass-loss rate were derived for all the non-detections with the exception of GJ\,1214\,b and WASP-127\,b. In the case of the detections, we found a constant day-to-night side zonal wind for the three planets with hot thermospheres but a relatively low mass-loss rate, which is consistent with previous findings. While HD\,189733\,b is confirmed to have a shallow metastable helium atmosphere, HAT-P-11\,b, and WASP-69\,b tend to have a more extended thermosphere with possibly a small exosphere.\\
The correlation between $\delta$R$_p$/H and M$_{\star}$ confirms that planets around K dwarfs are favored to have metastable helium in their atmosphere as proposed in \cite{oklopcic_helium_2019}. The correlation between $\delta$R$_p$/H and XUV flux proposed and described in several studies also seems to remain a good indicator, but is much more model-dependent that with the stellar mass, and thus requires caution when comparing studies. We also point out that the EUV emission is linked to the coronal heavy element abundance and thus to the stellar metallicity \citep{poppenhaeger_helium_2022}. The use of the $\dot{m}$ instead of $\delta$R$_p$/H should be utilized more as it is physically more representative of exoplanet thermospheres, although it remains ambiguous to draw a strong conclusion at the population level for now. This  could be improved by gathering higher quality datasets even for non-detections. Future studies and proposals could use as guidelines the $\delta$R$_p$/H versus M$_{\star}$ correlation to build robust science cases as it is the less model-dependent correlation but one should not forget that studying planets outside this soft spot can turn out to be as important.\\
Finally, we want to draw attention to the necessity of building reproducible observations from the proposal phase taking into account weather losses to get robust results. This lack of reproducibility for helium studies is frequent and calls for more than one transit observation per target. In this context, the NIRPS consortium will observe over 5 years more than 75 gas-dominated planets with at least two transits for each of them as part of its Guaranteed Time Observations (GTO) program. It will, therefore, provide an extended sample to study exoplanet atmospheres as a population, unlocking constraints on the origin of the Neptunian desert and planet evolution.

\begin{acknowledgements}
Based on observations obtained at the Canada-France-Hawaii Telescope (CFHT) which is operated from the summit of Maunakea by the National Research Council of Canada, the Institut National des Sciences de l'Univers of the Centre National de la Recherche Scientifique of France, and the University of Hawaii. The observations at the Canada-France-Hawaii Telescope were performed with care and respect from the summit of Maunakea which is a significant cultural and historic site. We thank the anonymous referee for their comments that improved the overall quality of this work. R. A. is a Trottier Postdoctoral Fellow and acknowledges support from the Trottier Family Foundation. This work was supported in part through a grant from the Fonds de Recherche du Qu\'ebec - Nature et Technologies (FRQNT). This work was funding by the Institut Trottier de Recherche sur les Exoplan\`etes (iREx). D. L., R. D., M. R. would like to acknowledge funding from the National Sciences and Research Council of Canada (NSERC). This project has received funding from the European Research Council (ERC) under the European Union's Horizon 2020 research and innovation program (project {\sc Spice Dune}, grant agreement No 947634; grant agreement No 730890, project SACCRED, grant agreement No 716155, project ASTROFLOW, grant agreement No 817540).  J.D.T was supported for this work by NASA through the NASA Hubble Fellowship grant $\#$HST-HF2-51495.001-A awarded by the Space Telescope Science Institute, which is operated by the Association of Universities for Research in Astronomy, Incorporated, under NASA contract NAS5-26555. J. F. D. acknowledges funding from the European Research Council (ERC) under the H2020 research \& innovation programme (grant agreement $\#$740651 NewWorlds). We acknowledge funding from the French ANR under contract number ANR\-18\-CE31\-0019 (SPlaSH). A.C. and X.D. acknowledge funding from the Investissements d'Avenir program (ANR-15-IDEX-02), through the funding of the "Origin of Life" project of the Grenoble-Alpes University.
\end{acknowledgements}



\bibliographystyle{aa}
\bibliography{Helium_SPIRou} 



\appendix

\begin{appendix}
\section{Master out spectra}
\begin{figure*}
\includegraphics[width=\textwidth]{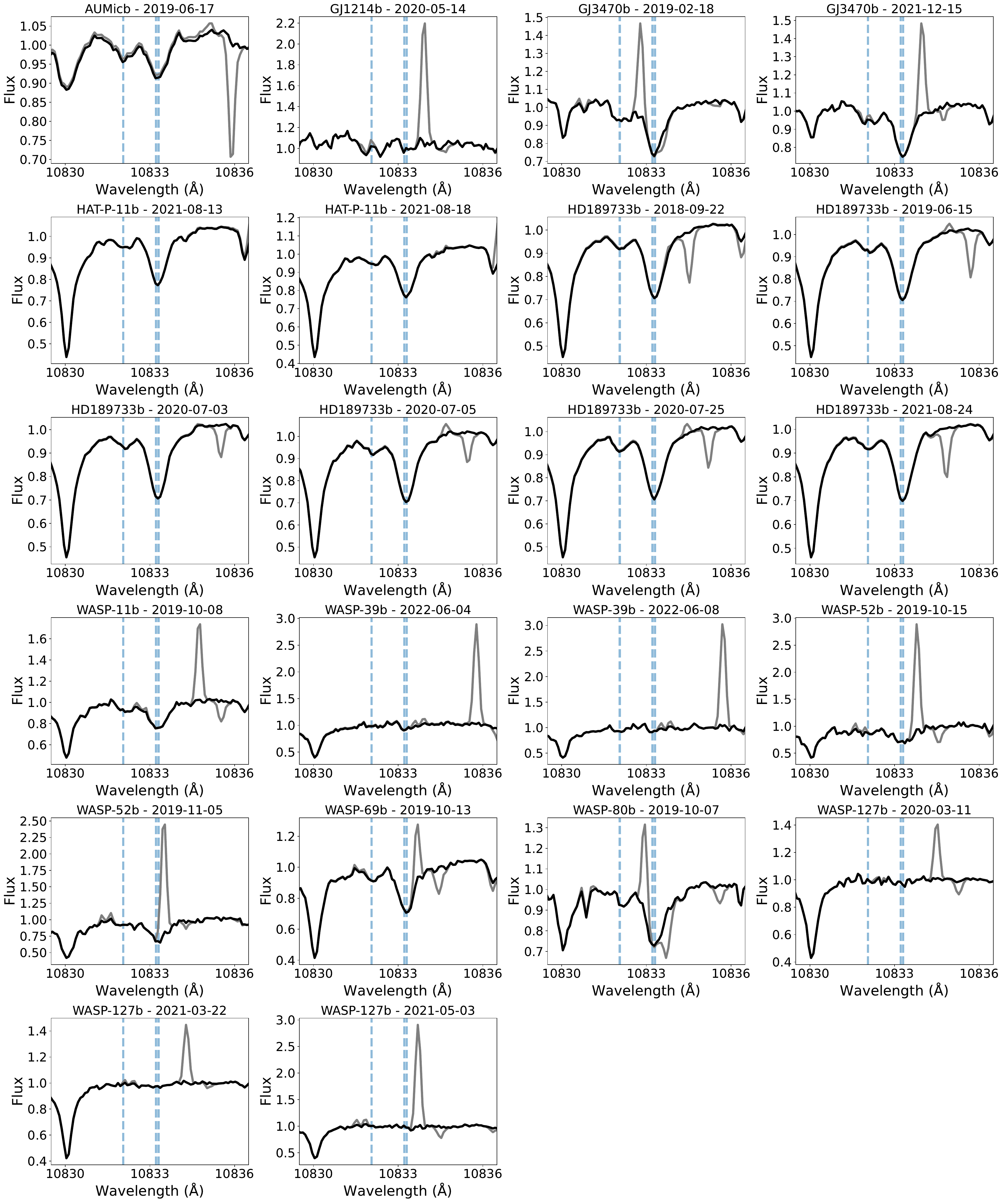}
\caption{Master out spectra before (grey) and after (black) telluric correction for each time series of the eleven planets. The vertical blue dashed lines indicate the helium lines' positions.}
    \label{fig:Master_out}
\end{figure*}

\section{Transmission spectroscopy map}

\begin{figure*}
\includegraphics[width=\textwidth]{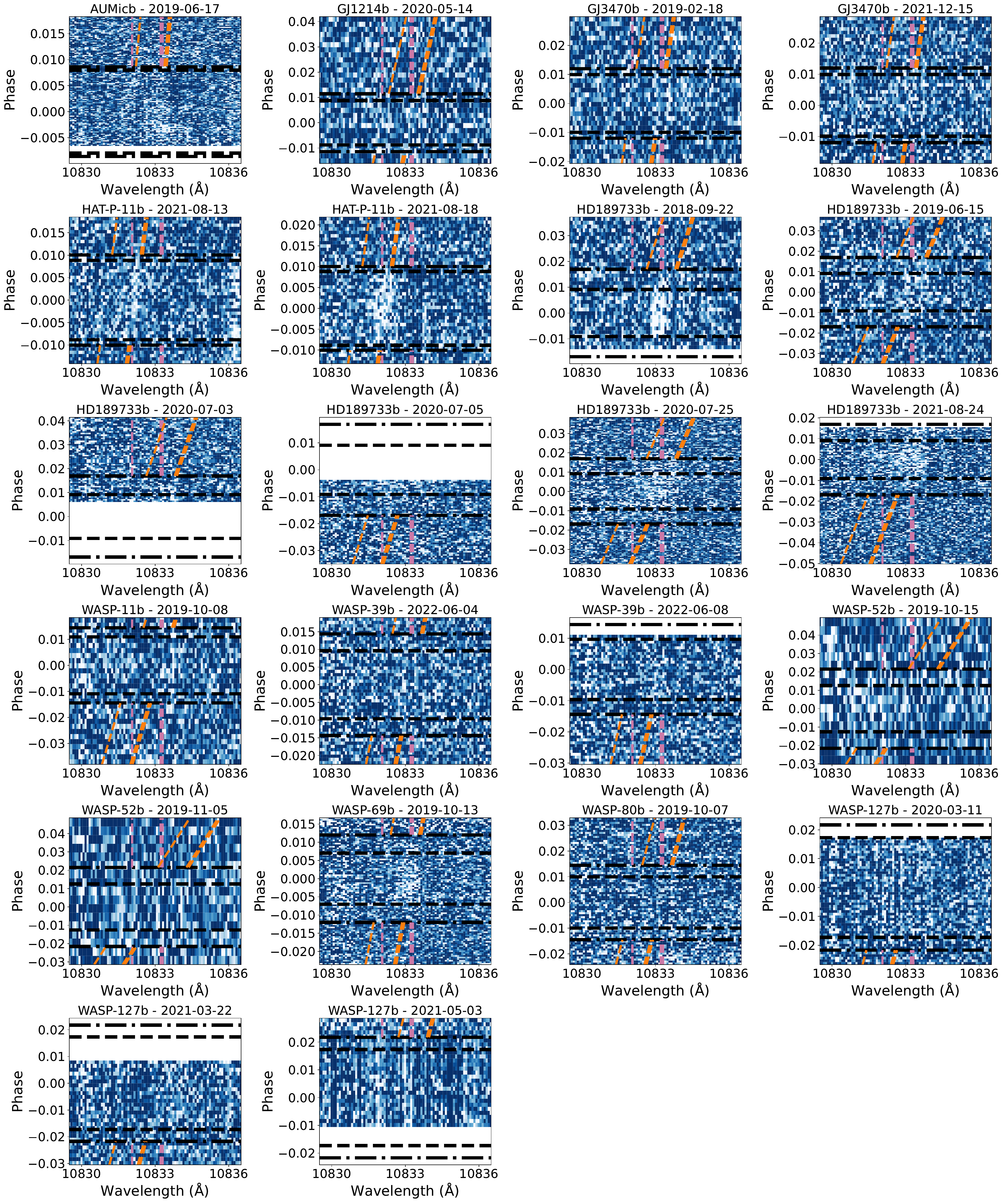}
\caption{Transmission spectroscopy maps in the star rest frame for each transit of the eleven planets. The dark dashed horizontal lines are the contact lines from bottom to top: t$_1$, t$_2$, t$_3$, and t$_4$. The orange dashed lines indicate the helium lines' positions in the planet rest frame. The vertical pink dashed lines indicate the helium lines' positions in the star rest frame.}
    \label{fig:TS_map_night}
\end{figure*}

\section{Red noise contribution}

\begin{figure*}
\includegraphics[width=\textwidth]{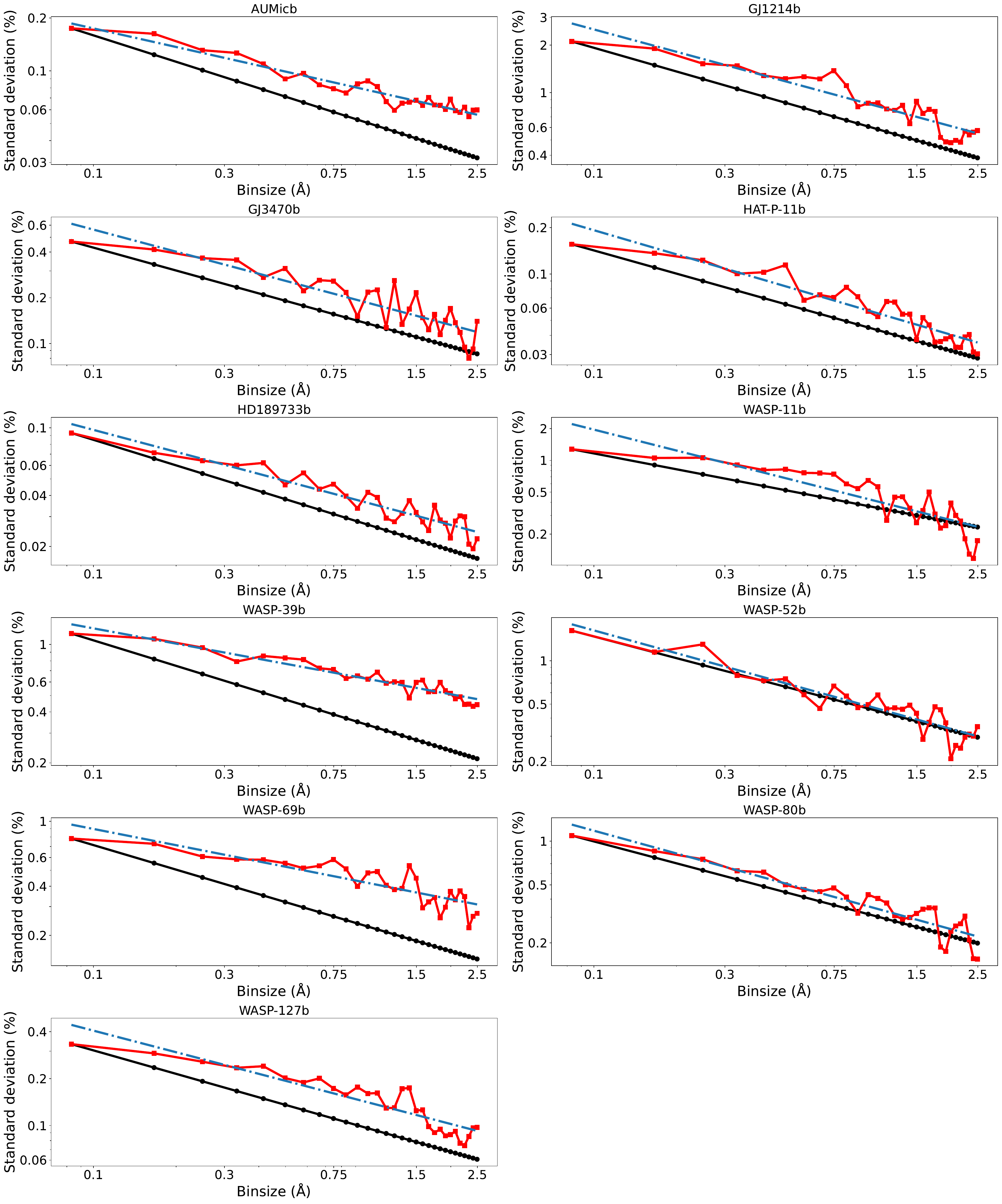}
\caption{Allan plot estimated on the average transmission spectra of the eleven planets. The dark dotted lines correspond to the white noise as function of the bin size. The white noise on 1 original pixel is estimated as the standard deviation of the transmission spectrum. The red dotted curves are the standard deviation of the transmission spectrum after various binning size. The dashed blue lines are the best fit applied of the reds curves.}
    \label{fig:Allan}
\end{figure*}

\section{Bootstrap simulation}

\begin{figure*}
\includegraphics[width=\textwidth]{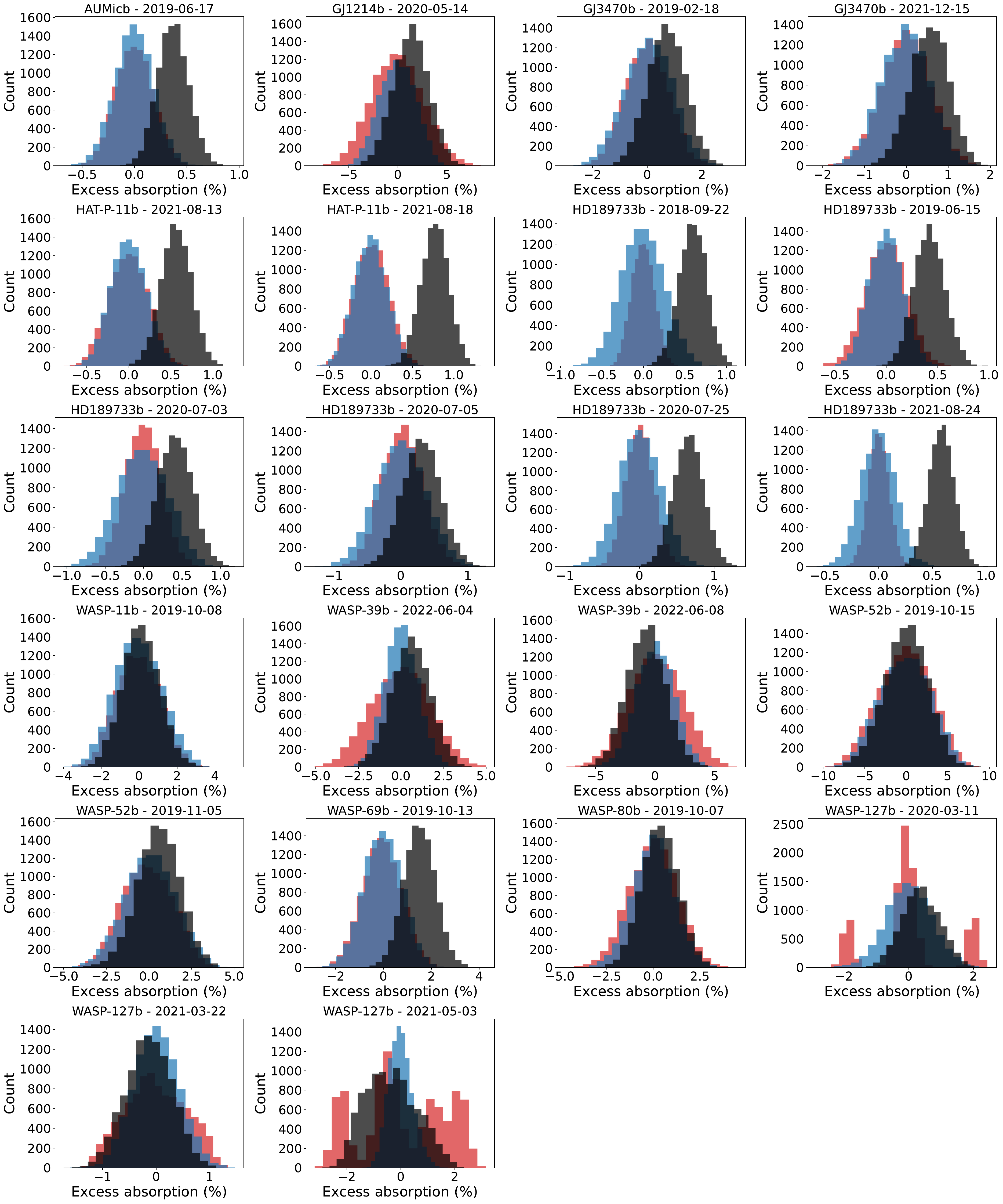}
\caption{Bootstrap analysis for each transit of the eleven planets in the star rest frame. The red, blue, and black distributions correspond to the out-out, in-in, and in-out random scenarios with 10000 iterations.}
    \label{fig:Boot_strap}
\end{figure*}

\end{appendix}

\end{document}